\documentclass[11pt, aas_macros]{article} % use larger type; default would be 10pt
\pdfoutput=1

\usepackage[utf8]{inputenc} % set input encoding (not needed with XeLaTeX)
\usepackage{jcappub}
\usepackage{rotating} %rotate table headings
\usepackage{multirow}
\usepackage{graphicx} % support the \includegraphics command and options
\usepackage{subfig}
\usepackage{booktabs} % for much better looking tables
\usepackage{array} % for better arrays (eg matrices) in maths
\usepackage{paralist} % very flexible & customisable lists (eg. enumerate/itemize, etc.)
\usepackage{verbatim} % adds environment for commenting out blocks of text & for better verbatim
\usepackage{subfig} % make it possible to include more than one captioned figure/table in a single float
\usepackage{epsfig}
\usepackage{amsmath}

\usepackage{fancyhdr} % This should be set AFTER setting up the page geometry
\usepackage{sectsty}
\allsectionsfont{\sffamily\mdseries\upshape} % (See the fntguide.pdf for font help)
\usepackage[nottoc,notlof,notlot]{tocbibind} % Put the bibliography in the ToC
\usepackage[titles,subfigure]{tocloft} % Alter the style of the Table of Contents

\newlength{\figlength}
\usepackage{soul}

\usepackage{hyperref}

\newcommand{\adsurl}[1]{\href{#1}{ADS}} 
\providecommand{\url}[1]{\href{#1}{#1}}

\title{Simulations for a next-generation UHECR observatory}
\author[a,b]{Foteini Oikonomou,}
\author[c,a]{Kumiko Kotera,}
\author[a,d]{Filipe B. Abdalla}
\affiliation[a]{Astrophysics Group, Department of Physics and Astronomy, University College London,\\
Gower Street, London WC1E 6BT, United Kingdom.}
\affiliation[b]{Department of Physics, Pennsylvania State University, University Park, PA 16802,USA.}
\affiliation[c]{Institut d'Astrophysique de Paris
UMR7095 -- CNRS, Universit\'e Pierre \& Marie Curie,
98 bis boulevard Arago
F-75014 Paris, France.}
\affiliation[d]{Department of Physics and Electronics, Rhodes University, PO Box 94, Grahamstown, 6140, South Africa.}

\emailAdd{oikonomou@psu.edu}
\emailAdd{kotera@iap.fr}
\emailAdd{fba@star.ucl.ac.uk}

\abstract{We explore the potential of a future, ultra-high energy cosmic ray (UHECR) experiment, that is able to overcome the limitation of low statistics, to detect anisotropy in the arrival directions of UHECRs. We concentrate on the lower energy range of future instruments ($E\gtrsim 50\,$EeV), where, if the UHECR source number density is not too low, the sources should be numerous enough to imprint a clustering pattern in the sky, and thus possibly in the UHECR arrival directions. Under these limits, the anisotropy signal should be dominated by the clustering of astrophysical sources {\it per se} in the large-scale structures, and not the clustering of events around individual sources. We study the potential for a statistical discrimination between different astrophysical models which we parametrise by the number density of UHECR sources, the possible bias of the UHECR accelerators with respect to the galaxy distribution, and the unknown fraction of UHECRs that have been deflected by large angles. We demonstrate that an order-of-magnitude increase in statistics would allow to discriminate between a variety of astrophysical models, provided that a sub-sample of light elements can be extracted, and that it represents a fraction $\gtrsim 70\%$ of the overall flux, sensitive to the UHECR source number density. Discrimination should be possible even without knowledge of the composition of the UHECRs, as long as the data are proton-dominated. We find that an anisotropy at the $99.7\%$ level should be detectable when the number of detected events exceeds 2000 beyond 50 EeV, as long as the composition is proton dominated, and the number density of UHECR sources is relatively high, $\bar{n} \geq 10^{-3}~{\rm Mpc}^{-3}$. If the UHECR sources are strongly biased relative to the galaxy distribution, as are for example galaxy clusters, in which the sources might be embedded, an anisotropy at the $99.7\%$ level should be detectable once the number of detected events exceeds 1000, if the fraction of protons at the highest energies is $\gtrsim 60\%$.}

\keywords{}

\arxivnumber{}

\date{\today} 

\begin{document}
\maketitle
\flushbottom

\section{Introduction}
\label{sec:intro}

The origin of ultra-high energy cosmic rays (UHECRs) remains unknown despite decades of experimental and theoretical efforts (see \cite{NW00,Bhatta00,KO11,Letessier11} for reviews). The propagation distance of UHECRs is limited to a few hundred Mpc at the highest energies due to energy losses via interactions with the Cosmic Microwave Background, that induce the so-called GZK energy cut-off \cite{G66,Zatsepin-Kuzmin}. The matter distribution is not homogeneous over such distances, hence if UHECRs are extragalactic, one expects an anisotropy in their arrival direction distribution, reflecting the inhomogeneity of the source distribution, if magnetic deflections do not completely smear their trajectories.

The observed hints for a departure from isotropy at energies beyond the GZK cutoff, remain insufficient to draw conclusions as to the sources of UHECRs with available data (e.g. \cite{kashti2008,Abreu10,Oikonomou13}). As a result of the absence of a clear correlation signal with luminous sources, and of the recent experimental evidence for an increasingly heavy UHECR composition at the highest energies \cite{2010PhRvL.104i1101A,2013arXiv1307.5059T}, the prospects for source identification with a next-generation detector are debated.

The current generation of UHECR detectors, the Pierre Auger Observatory (hereafter Auger) \cite{2004NIMPA.523...50A} and the Telescope Array (TA) \cite{AbuZayyad201287,Tokuno201254}, collect super-GZK events (at energy $E\gtrsim 60\,$EeV) at a rate $\lesssim 2$~month. The next UHECR experiment could be a space-based telescope such as the proposed JEM-EUSO, which is proposed to be mounted on the International Space Station in $\sim 2020$. If launched, it will survey the night sky for the ultra-violet fluorescence and Cherenkov radiation produced when a UHECR hits the Earth's atmosphere \cite{Adams:2012tt}, with a near uniform exposure over the full sky. Depending on the operation mode, JEM-EUSO is expected to reach $9-20$ times the annual Auger exposure at $100$~EeV. In {\it nadir} mode it will be sensitive to UHECRs with energy $E \geq 40$~EeV and fully efficient beyond $60-70$~EeV.

A number of simulation studies have been performed with the aim of assessing the anisotropy discovery potential of a next-generation, full sky observatory. Studies so far have mostly focused on the anisotropy expected at the highest energies $E\gtrsim 100\,$EeV, where the energy losses are such that the source horizon is considerably reduced. If the source density is low ($ \lesssim 10^{-5}~{\rm Mpc}^{-3}$), only a handful of objects should be contributing to the total observed flux as demonstrated in~\cite{Blaksley13}. In this regime, the anisotropy signal is dominated by the clustering of events around these few sources. Reference~\cite{Rouille14} showed that even in very unfavourable composition scenarios (with, e.g., no protons accelerated to the highest energies), a next-generation JEM-EUSO type detector, should allow the measurement of a significant anisotropy signal, assuming the sources follow the spatial and luminosity distribution of the Two Micron Redshift Survey (2MRS) galaxies \cite{2MRS}. 

In reference~\cite{Denton:2014nga}, the sensitivity of JEM-EUSO to large-scale anisotropies in the UHECR arrival distribution, using a spherical harmonic analysis, was estimated. It was demonstrated that JEM-EUSO would be able to detect any such large-scale anisotropy, even if it much weaker than current experiments are sensitive to. In reference~\cite{Decerprit12}, methods for constraining the source density and overall particle deflection angles with current and future observatory statistics were developed. 

In this work, we study the sensitivity of a next-generation UHECR detector to the UHECR anisotropy signal expected for large source numbers. Instead of focusing on the high energy end, where the number of sources is reduced to a few, we concentrate on the lower energy range of the future instruments ($E\gtrsim 50\,$EeV), where the sources are numerous enough to imprint a clustering pattern in the sky and thus possibly in the particle arrival directions. Under these limits, the anisotropy signal should be dominated by the clustering of astrophysical sources {\it per se} in the large-scale structures, and not the clustering of events around individual sources. Although the UHECR source density is unknown, a relatively high source number density, $\bar{n}$, is favoured by the observed clustering of UHECRs, whereas models with $\bar{n} < 10^{-5}~{\rm Mpc}^{-3}$ are strongly disfavoured \cite{Abreu:2013kif}.
 
One of the properties of UHECR sources that can be used to constrain their spatial distribution, is their relative bias with respect to that of galaxies. The authors of \cite{WFP96,kashti2008} studied whether one can constrain such a bias through a statistical analysis of UHECR arrival directions. In the case of transient UHECR sources, and a proton dominated composition, a different bias (degenerate with the former) is expected to occur~\cite{KL08b, Kalli10}. The conclusions of these studies, that such a bias should be detectable, motivate us to take a closer look at the potential of a future UHECR experiment to detect such an effect.

The latest measurements of Auger, indicate that at super-GZK energies the data are not compatible with a pure proton composition \cite{2010PhRvL.104i1101A,2013arXiv1307.5059T}. The TA results show the same trend within the systematics \cite{Abbasi14,Pierog13}. Heavier elements, that would be deflected by intergalactic magnetic fields, would dilute any potential correlation with the distribution of the sources in the sky. In the framework of a future UHECR experiment, capable of distinguishing light from heavy showers, and thus able to extract a sub-sample of protons, one can estimate the fraction of protons in the sample, required in order to detect a significant anisotropy signal, and discriminate between different source distribution models. Another option is to conservatively model the contribution of the heavy nuclei as an isotropic background, and determine the fraction of protons necessary to obtain a significant signal when taking into account all the observed particles. 

We perform Monte Carlo modelling of the unknown UHECR sources and forecast the expected UHECR arrival distributions in a range of astrophysical models. We focus on the prospect of a statistical discrimination between different astrophysical scenarios and perform a scan over the allowed parameter space, which includes the unknown UHECR source density, the unknown fraction of protons at the highest cosmic ray energies, and the possible bias of the UHECR source distribution with respect to the galaxy distribution for which we consider a wider range of physically motivated models than previous studies. We present the probability of ruling out individual models for the sources of UHECRs, for the number of events expected to be detected within a few years of operation of a detector with an annual exposure comparable to that expected for JEM-EUSO. The relative merit of a next-generation experiment, that focuses on high exposure, over a ground based experiment, that can perform more precise measurements at the cost of more modest statistics, is an important question. We demonstrate here that an order-of-magnitude increase in statistics would allow to discriminate between a large variety of astrophysical models, provided that a sub-sample of light elements can be extracted, and that it represents a fraction $\gtrsim 30\%$ of the overall flux. A significant anisotropy is also expected, even if the composition cannot be determined, as long as a significant fraction of protons is present in the dataset.

Our results are general and apply to any future UHECR detector with an order of magnitude higher exposure than current experiments. Given the prospect of JEM-EUSO being the next experiment dedicated to UHECR detection, we assume some of its expected characteristics, namely its expected annual exposure, detection sensitivity as a function of energy, nearly uniform full sky exposure, pointing, and energy resolution \cite{Adams13}. Throughout we take into account the energy losses of UHECRs as they propagate through the background photon fields, and their deflections in intervening magnetic fields.
 
\section{Bias prescription of UHECR source clustering}
\label{sec:bias}

It is well established observationally that different galaxies are biased tracers of the underlying mass distribution in the Universe, and that different galaxy types cluster to the mass distribution with varying strengths (see \cite{Kaiser84,ST99} for possible theoretical explanations). Clusters of galaxies cluster more strongly than galaxies themselves, whereas some subtypes of galaxies, in particular luminous red galaxies and AGN, are observed to cluster more strongly than the average galaxy field. Observationally, a way to constrain the sources of UHECRs is to consider their relative bias to that of galaxies overall. From this point on, in this work, with the term bias we refer to the bias of UHECR sources relative to catalogued galaxies, and not the bias relative to the underlying mass distribution.

Denoting the galaxy density field $\rho_{\rm g}$ and its mean value $\bar{\rho_{\rm g}}$ we can write the local galaxy fractional overdensity as $\delta_{\rm g} = \rho_{\rm g}/\bar{\rho_{\rm g}} - 1$. Similarly, we can express the unknown local fractional overdensity of UHECR sources as $\delta_{\rm s} = \rho_{\rm s}/\bar{\rho_{\rm s}} - 1$. The simplest, often assumed, relation between two overdensity fields, in our case $\delta_{\rm s}$ and $\delta_{g}$, is a linear bias of the form: 
\begin{equation}
\delta_{\rm s} = b \cdot \delta_{\rm g}.
\end{equation}
This, widely used model, cannot hold in all cases. For example, it can result in negative densities if $b > 1$. It is however generally a good approximation on cosmological scales, where the density fluctuations $\delta \ll 1$.

Here, we consider a number of toy models for the bias of UHECR sources, with the aim of capturing different astrophysical scenarios for the origin of UHECRs. If UHECRs are accelerated in common sources, the overall UHECR distribution should follow that of galaxies. If, for example, UHECRs are accelerated primarily in young pulsars, then the distribution of UHECR sources should roughly follow the distribution of young starburst galaxies, which roughly follows that of ordinary galaxies (e.g., \cite{Owers2007}). On the other hand, if UHECRs are accelerated in uncommon, extreme sources such as AGN and radio galaxies, which tend to be found in over-dense regions, then the distribution of UHECR sources should be more strongly clustered than that of galaxies. With the aim of bracketing the above scenarios for the bias $b$ of the unknown UHECR sources with respect to that of galaxies, we consider the following models:
\begin{itemize}
\item An {\it isotropic} model (I), where $\delta_{\rm s} = 0$ everywhere.
\item An {\it unbiased} model (UB), where $\delta_{\rm s} = b \cdot \delta_{\rm g}$, with $b = 1$. The unbiased model describes a situation in which all galaxies (or sources in these galaxies) are equally likely to accelerate UHECRs to the highest energies. 
\item A {\it linear} bias model (L), where $\delta_{\rm s} = b \cdot \delta_{\rm g}$, with $b = 3$. This model better describes a scenario in which observed UHECRs originate in sources that tend to preferentially populate overdense regions, such as AGN or radio galaxies (e.g., \cite{SP89}).
\item A {\it threshold} bias model (TH), where $\delta_{\rm s} = \delta_{\rm g}$ if $\delta_{\rm g} > \delta_{\rm min}$ else $\delta_{\rm s} = -1$. This is a more extreme, ad hoc version of the linear bias model, in which only the densest regions (i.e. primarily galaxy clusters) contain sources responsible for the production of observed UHECRs.
\end{itemize}
In section \ref{sec:maps}, we show how these models are applied to the 2MRS catalogue to obtain models of the expected UHECR intensity maps.

\section{UHECR intensity maps and simulations}
\label{sec:maps}
\subsection{Source distribution}

We use the $K > 11.75$ 2MRS catalogue, as the basis of our model of the expected UHECR source distribution. For simplicity, we consider identical UHECR sources, which produce a power-law spectrum with maximum acceleration energy $E_{\rm max}=10^{21}$~eV and injection spectral index $\alpha=-2.0$. We weight the 2MRS galaxies by the inverse of the selection function, using the method presented in \cite{Davis10}, to correct for the incompleteness of the magnitude limited survey. We consider recession velocities in the Local Group frame to correct, to some extent, for the peculiar velocities of nearby galaxies (see discussion in e.g., \cite{Erdogdu06,Oikonomou13}). 

We pixelise the sky using the HealPix package \cite{2005ApJ...622..759G}. To construct models of the UHECR source distribution, we split the galaxy distribution into radial shells of equal thickness. We have chosen a thickness of 11 Mpc as a compromise between degrading the resolution of the analysis and having large enough shells to capture large structures (as a reminder the typical radius of a galaxy cluster is a few Mpc). We estimate the fractional overdensity of sources in each cell, $i$, by $\delta_{{\rm s},i} = \rho_{{\rm s},i}/\bar{\rho}_{{\rm s},i} -1$, where the average, $\bar{\rho}_{{\rm s},i}$, is obtained by averaging over all cells in each of the radial shells defined. We compute the value of the bias, $b$, for each individual radial shell. Consequently we apply the bias to each cell $i$ in that shell. In the {\it linear} bias model, in some underpopulated bins we end up with negative densities for the reason explained in section \ref{sec:bias}. Given that $\delta_{\rm s} < -1$ is unphysical we naturally need to set some under-dense regions to have a number density equal to zero. This, unavoidable procedure, does not make a big difference to the expected correlation signal in practice, as the cross-correlation mainly picks out the high density regions.

To construct the {\it threshold} bias model, we use the procedure outlined below. A regular hard thresholding procedure would usually involve setting all the pixels below a fixed $\delta_{min}$ to $\delta_{\rm s} = -1$. However, given that the galaxy field changes as a function of distance (number density of galaxies in the sample increases and then decreases due to selection effects), it would not be possible to maintain a fixed threshold for this situation. Instead, we choose to threshold via a proportion of pixels, setting the cooler pixels to $\delta_{\rm s} = -1$. We keep the fraction of volume retained in the threshold model constant as a function of redshift. We determine the proportion of hot pixels to be retained in the model by calibrating it against the 2MRS cluster catalogue of \cite{Crook07}. In practice, this means approximately $30\%$ of all 2MRS galaxies contribute to the UHECR flux in the {\it threshold} bias model and this fraction is roughly independent of redshift. The cluster catalogue of \cite{Crook07} itself could have served as a model for a UHECR source distribution that follows that of massive groups and clusters, but it suffers from the disadvantage that is stops at $z = 0.034$.

We weight each radial shell for its expected contribution to the arriving UHECR flux following
\begin{equation}
\label{eq:weight}
\omega(d_{{\rm L}})_{{\rm flux}} = \frac{1}{{d_{{\rm L}}}^2} \int^{E_{{\rm i, max}}} _{E_{{\rm f}}'} {\rm d}E_{{\rm i}} \int^{E_{{\rm i}}} _{E_{{\rm f}}'} {\rm d}E_{{\rm f}} ~ \left| \frac{\partial P_p(d_{{\rm L}}, E_{{\rm i}}; E_{{\rm f}})}{\partial E_{{\rm f}}} \right|~I(E_{{\rm i}}),
\end{equation}
\noindent where $P_p(d_{{\rm L}}, E_{{\rm i}}; E_{{\rm f}})$ gives the probability of a proton arriving with energy above $E_{{\rm f}}$, if it was emitted with energy $E_{{\rm i}}$ by a source at luminosity distance $d_{L}$ \cite{2000ApJ...542..542B}. For $E_{{\rm i, max}}$, which is the maximum energy achievable through astrophysical processes, we take $10^{21}$~eV. Our results are insensitive to this choice, as UHECRs with energy beyond $10^{21}$~eV suffer rapid energy losses. For the intrinsic spectrum of UHECRs produced by UHECR sources, we consider a power law spectrum $I(E_{{\rm i}})=I_0~E_{{\rm i}}^\alpha~e^{-E_{{\rm i}}/E_{{\rm i, max}}}$, with index $\alpha = -2.0$ and normalisation $I_0$. For $d_{\rm L}$, we take the distance to the mean of each shell. 

For all UHECR energy loss calculations, we have used the results obtained by Monte Carlo in \cite{2001PhRvD..63b3002F, 2003JCAP...11..015F}\footnote{available at \url{http://www.desy.de/~uhecr/P_proton}}. We assume a homogeneous contribution to the expected UHECR flux by sources beyond 300 Mpc, where the 2MRS ends.

\begin{figure}[!th]
\centering
\includegraphics[width=0.32\columnwidth]{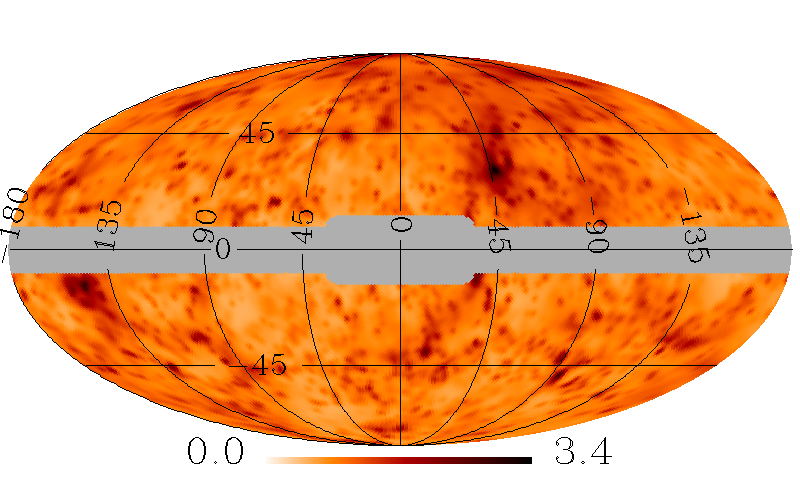}
\includegraphics[width=0.32\columnwidth]{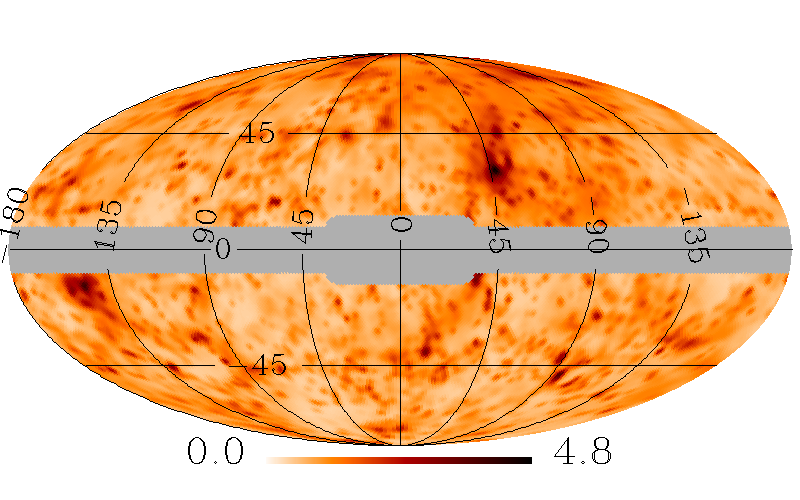}
\includegraphics[width=0.32\columnwidth]{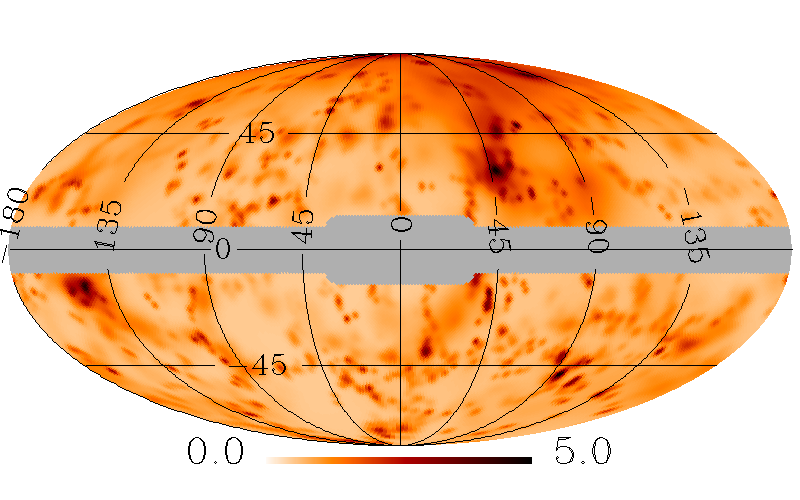}
\includegraphics[width=0.32\columnwidth]{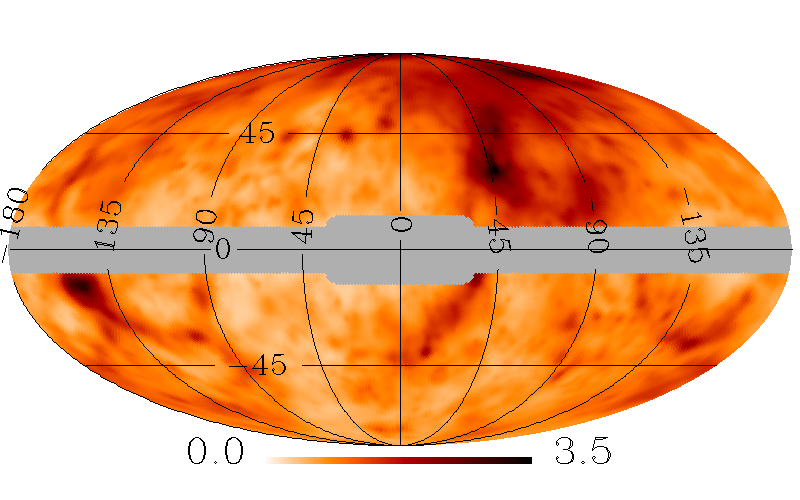}
\includegraphics[width=0.32\columnwidth]{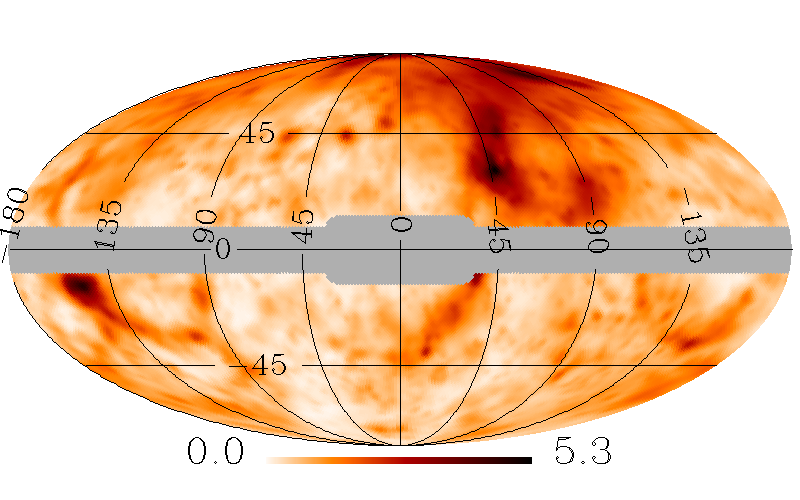}
\includegraphics[width=0.32\columnwidth]{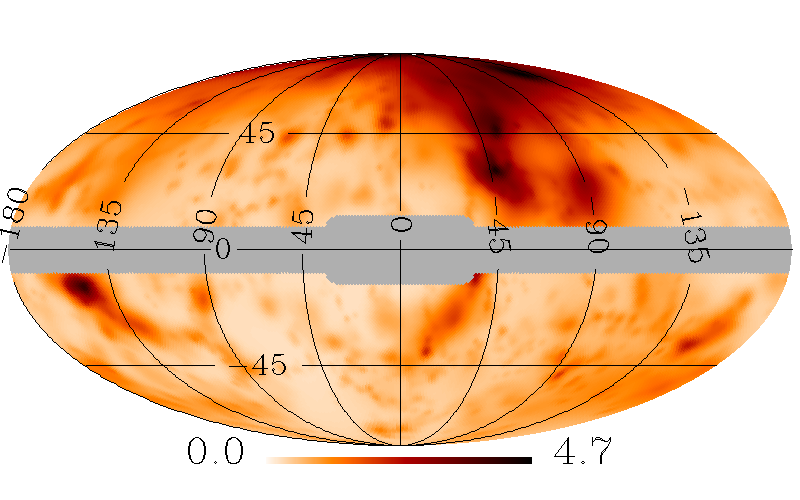}
\includegraphics[width=0.32\columnwidth]{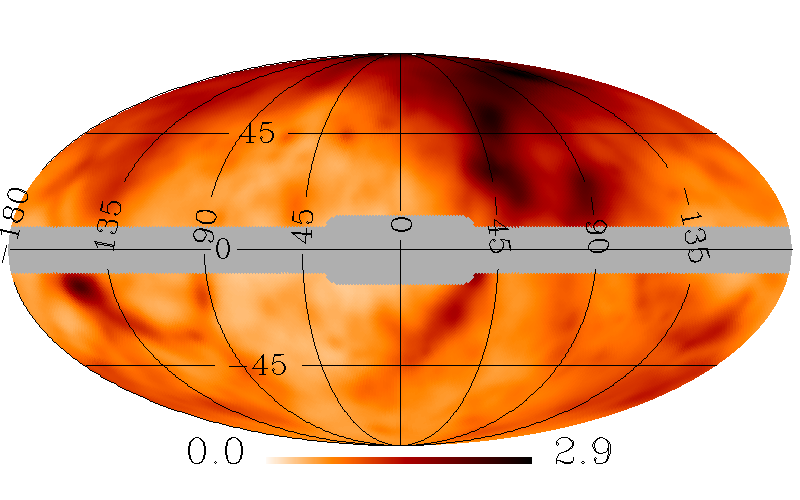}
\includegraphics[width=0.32\columnwidth]{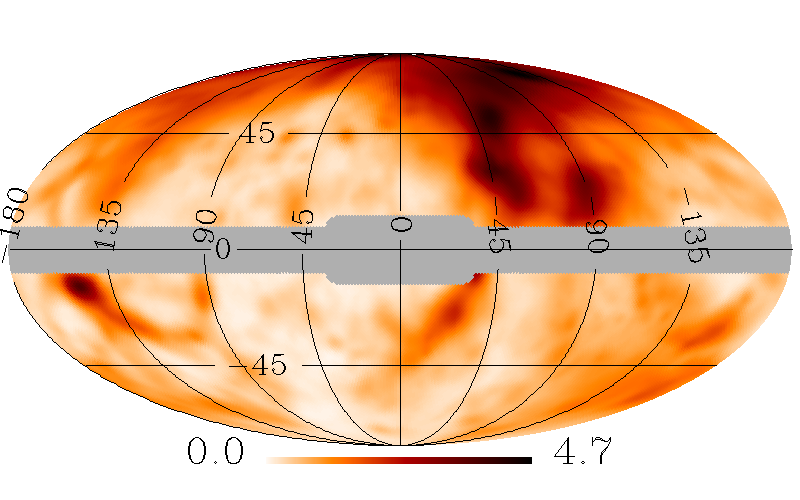}
\includegraphics[width=0.32\columnwidth]{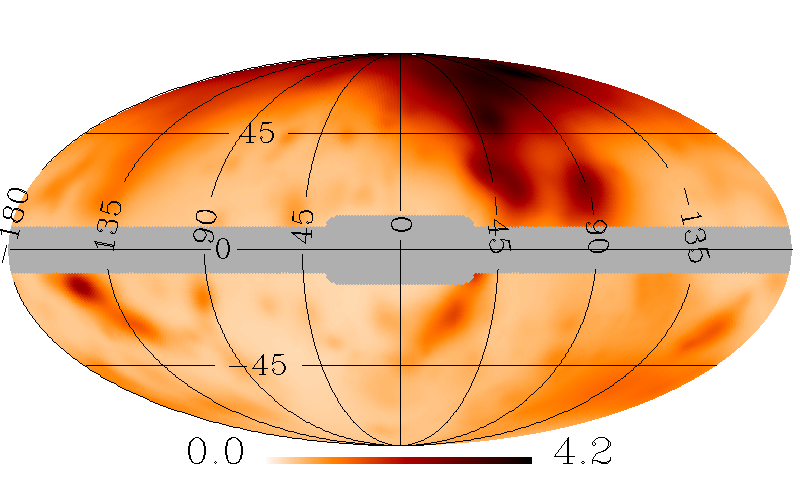}
\caption{The average expected UHECR intensity for UHECRs with energy above 50 EeV (top row), 80 EeV (middle), 100 EeV (bottom), in the {\it unbiased} source model (left panels), {\it linear} bias model (middle), {\it threshold} bias model (right), in Galactic coordinates, with $l=0^{\circ}$ at the centre of the map and $l$ increasing to the left.}
\label{fig:maps}
\end{figure}

We smooth the galaxy distribution with a Gaussian filter, with a redshift-independent amplitude, $\sigma = 350~{\rm km}~{\rm s}^{-1}$, following \cite{Davis10}. This, fixed smoothing length, is motivated by the dense sampling of the 2MRS. The smoothing length used corresponds to different physical lengths at different distances; the average resolution of the maps after smoothing is $\sim 5^{\circ}$. We mask the galaxy survey with the 2MRS mask which we dilate by $5^{\circ}$ (the average smoothing length) to remove the areas around the masked region that will have unavoidably been depleted by the smoothing procedure. In figure \ref{fig:maps}, we show the fluctuations in the mean of the average, expected UHECR intensity, for UHECRs with energy above 50, 80, 100 EeV, in the various bias models. As expected, the contrast in significantly enhanced in the linear and threshold bias models with respect to the unbiased model. The linear bias model has a higher contrast than the threshold bias model at 80 EeV and beyond. This is the result of using toy models with fixed parameters to model different astrophysical scenarios in the absence of statistically complete, observational samples. In any case, at these energies the UHECR horizon is small, making a clear distinction between the two models difficult in a statistical sense.

\subsection{Source density}

We generate a specific realisation of UHECRs by drawing the expected number of sources contributing one or more observed cosmic rays from the smoothed underlying UHECR source distribution in our models. The total number of sources depends on the number density of UHECR sources, which we determine as follows: defining the effective volume to be contained within the radius $r_{\rm GZK}$ at which the surviving fraction of UHECRs drops to $e^{-1}$, we draw from the smooth maps $N$ sources, such that the source number density, $\bar{n} = N/V_{\rm GZK}$, takes the value we have chosen, with $V_{\rm GZK} = (4/3) \pi r_{\rm GZK}^{3}$. The effective GZK radius, $r_{\rm GZK}$, depends on the UHECR energy. For the source number density, we consider values in the range $\bar{n} = 10^{-4}~{\rm Mpc}^{-3}-10^{-2}~{\rm Mpc}^{-3}$, the former being close to the lower limit of the source density derived from UHECR clustering \cite{Abreu:2013kif}, and the latter the number density of bright galaxies.

\subsection{Treatment of UHECR deflections}

We account for the deflections expected to be suffered by UHECR protons at different energies, using an adhoc analytic scaling with energy, $\sigma^2 = 2^{\circ} \cdot 2^{\circ}+1.5^{\circ} \cdot 1.5^{\circ}(100~{\rm EeV}/E)^2$. This choice is aimed to reflect the expected angular resolution of JEM-EUSO of $ \lesssim 2^{\circ}$, and the energy dependent rigidity of proton UHECRs as they propagate through the extragalactic, and Galactic magnetic fields \cite{1996ApJ...472L..89W}.

\subsection{Energy resolution}

To account for the energy resolution of the detector, we conservatively assume a $30\%$ energy resolution. We implement this uncertainty as a Gaussian detector response for the reconstructed UHECR energy and perform our analysis with the reconstructed energies, instead of the actual energies injected in the simulations. We simulate events down to 10~EeV, i.e. $\sim 4~\sigma$ below the studied energy threshold of $50$~EeV.

\subsection{Expected number of events}

The absolute energy scale of the observed UHECR spectrum is not precisely known, due to the large systematic errors associated with the energy reconstruction of the primary UHECRs ($\sim 14\%$ in Auger, $\sim 21 \%$ in the TA). The UHECR spectra that have been presented by the TA and Auger appear to have a systematic disagreement, but can be brought into agreement by a $\sim 20\%$ rescaling of the energy (see \cite{spectrumWG2012}), well within the published systematic uncertainties. This introduces an uncertainty to the expected number of UHECRs above a certain energy for a future detector like JEM-EUSO. 

Here, for the expected number of events we use the estimate of \cite{Rouille14}. Assuming that either the Auger flux or the HiRes/TA flux hold over the whole sky, and fitting the measured spectra with simulations of expected UHECR spectra in various scenarios for the injected UHECR spectrum, the authors derived two fiducial spectra by averaging over simulated spectra in the various scenarios. The two fiducial spectra, one with the Auger absolute energy scale, and one with the TA absolute energy scale, were then used to determine the expected numbers of events for a detector with the detection efficiency presented in \cite{Adams13}, and an integrated exposure equal to $3 \times 10^{5}~{\rm km}^{2}~{\rm sr}~{\rm yr}$. The number of expected event was found to be 1100, 250 and 100 events above 50, 80, 100 EeV respectively for the Auger energy scale. For the TA absolute energy scale 2100, 580, and 260 events are expected for the same energy thresholds. Since in this work we exclude the masked region from the statistical analysis but fix the total number of expected events, the number of expected UHECRs in the field of view of the 2MRS models in a given trial is also determined by Monte Carlo and varies between realisations. 

\subsection{Statistical approach}

To quantify the expected sensitivity of JEM-EUSO to any anisotropy signal in the UHECR arrival directions, we use the statistic $X_{\rm M}$, introduced in \cite{kashti2008}:
\begin{equation}
X_{\rm M} = \sum_{i}\frac{\left( N_{{\rm CR}, i} - N_{{\rm iso}, i} \right) \cdot \left( N_{{\rm M}, i} - N_{{\rm iso}, i} \right)}{N_{{\rm iso}, i}}.
\label{eq:statistic}
\end{equation}
\noindent Here, $N_{{\rm CR}, i}$ is the number of UHECRs detected in bin $i$. $N_{{\rm iso},i}$ is the number of UHECRs expected to be detected in bin $i$ in the isotropic model and $N_{{\rm M},i}$ is the number of UHECRs expected to be detected in bin $i$ in model $M$ (i.e. either of the unbiased, linear and threshold models). We calculate $X_{\rm M}$ over $7^{\circ} \times 7^{\circ}$ angular bins to average over possible UHECR deflections. In practice, we have (arbitrarily) chosen to evaluate $X_{\rm M}$ assuming the UHECR source distribution follows the unbiased model so from here on we will write $X_{\rm UB}$. As a result of having previously smoothed the expected source distribution with a filter of a comparable smoothing length, $X_{\rm UB}$ does not vary significantly for bin sizes smaller than $\sim 7^{\circ} \times 7^{\circ}$.

\section{Results}

With the setup presented in sections \ref{sec:bias} and \ref{sec:maps} we calculate the anisotropy signal expected to be detected with five years of data with JEM-EUSO, assuming a total exposure of $3 \times 10^{5}~{\rm km}^{2}~{\rm sr}~{\rm yr}$. 

\subsection{Sensitivity to isotropic fraction of UHECRs}

In this section, to mimic the effect of a mixed/heavy UHECR composition, we show the sensitivity of JEM-EUSO to the expected anisotropy signal in the presence of a fraction of UHECRs that have been deflected by large angles. We assume that a fraction of UHECRs arrive without having retained a correlation with their sources, i.e. isotropically, and vary this fraction in our simulations. 

\begin{figure}[!ht]
\centering
\includegraphics[width=0.75 \figlength]{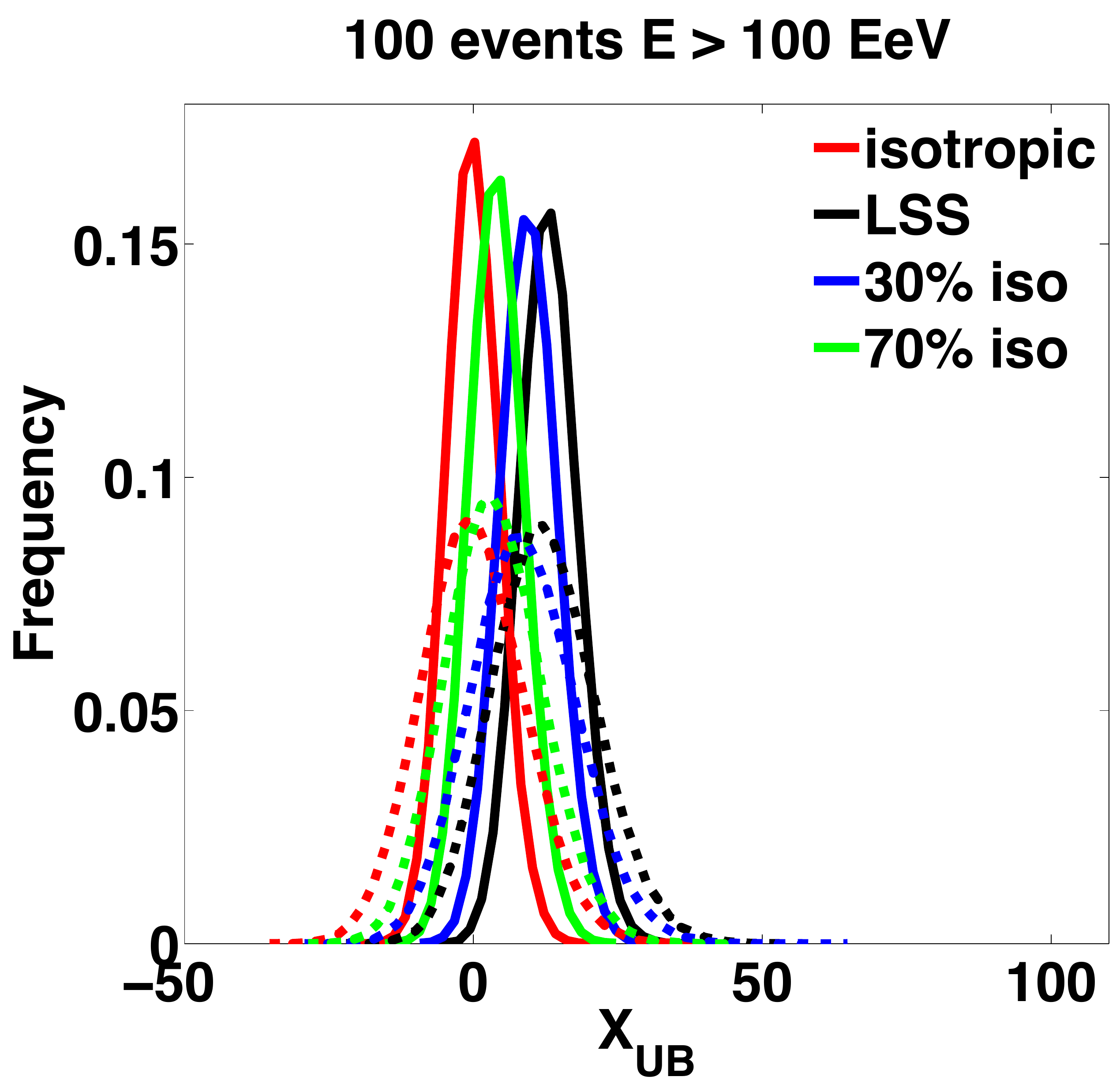}
\includegraphics[width=0.75 \figlength]{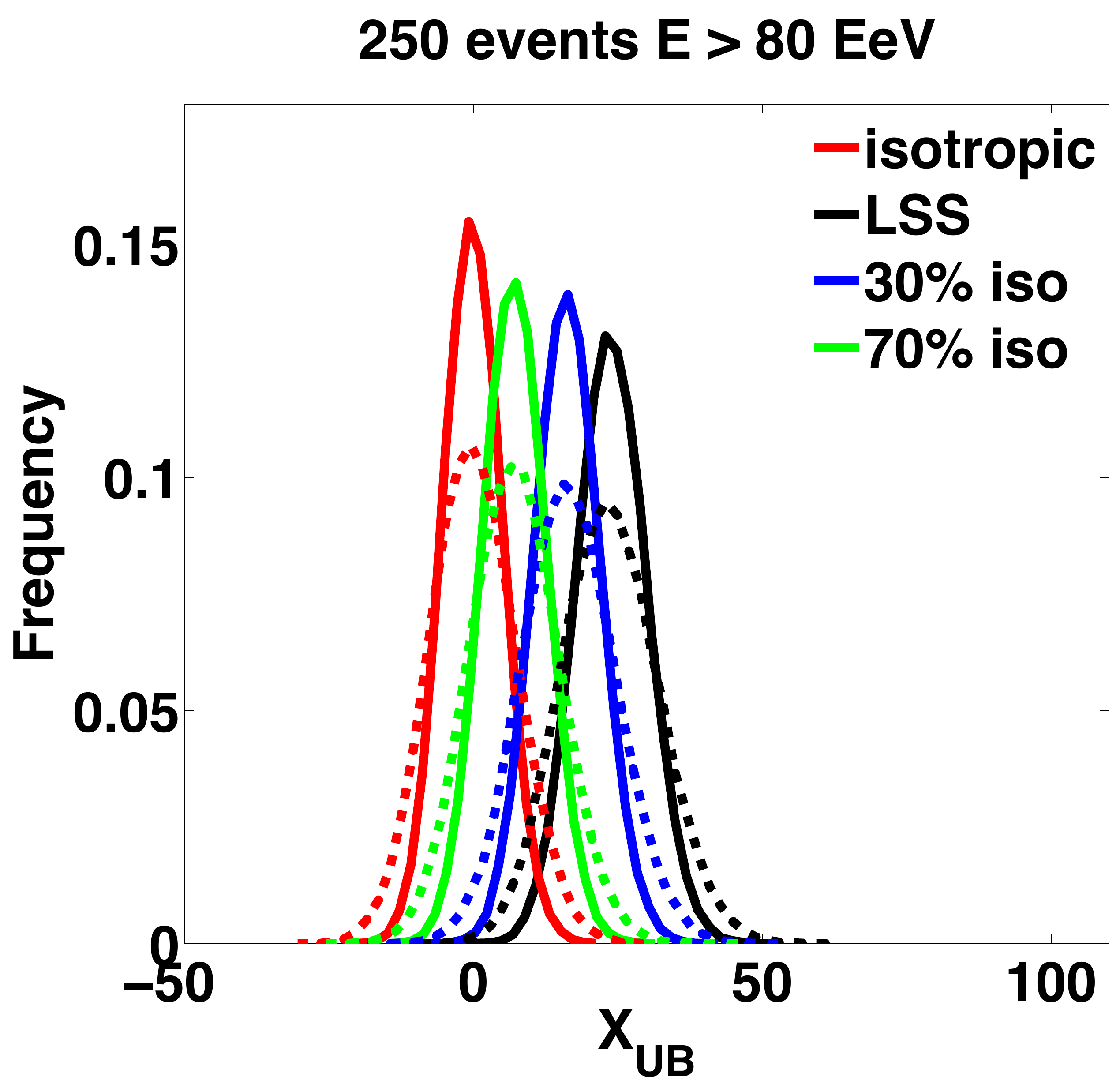}
\includegraphics[width=0.75 \figlength]{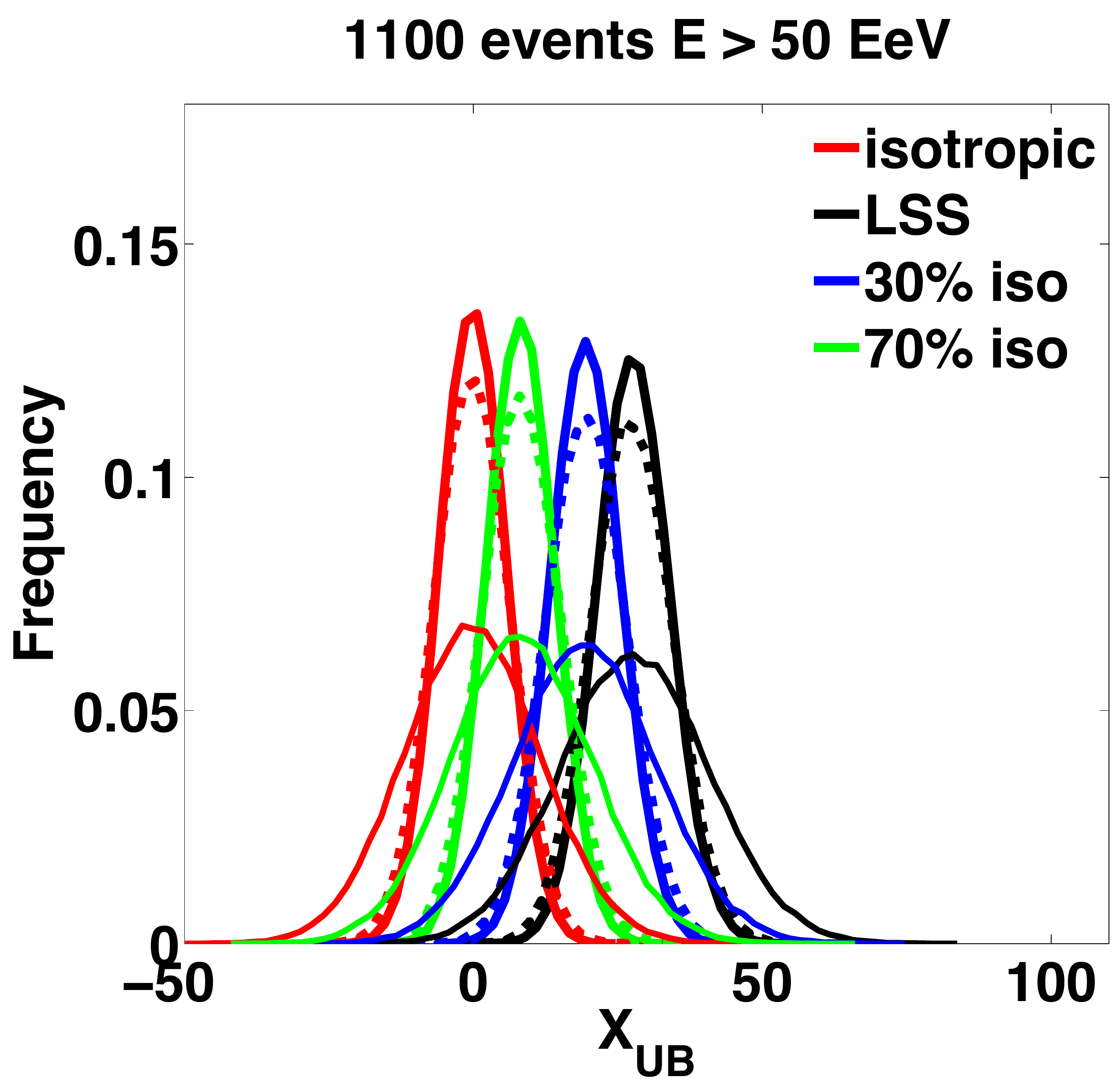}\\
\includegraphics[width=0.75 \figlength]{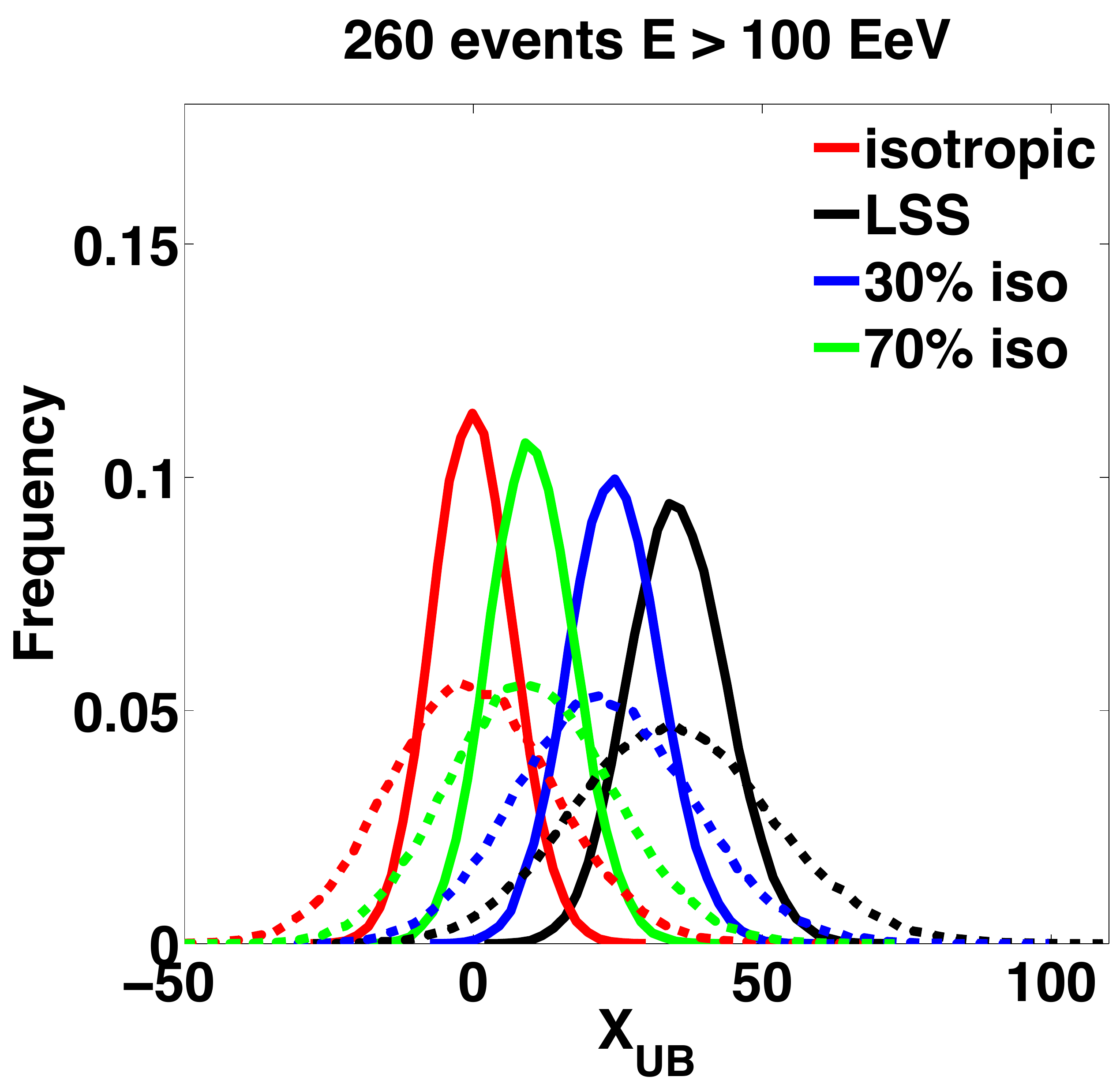}
\includegraphics[width=0.75 \figlength]{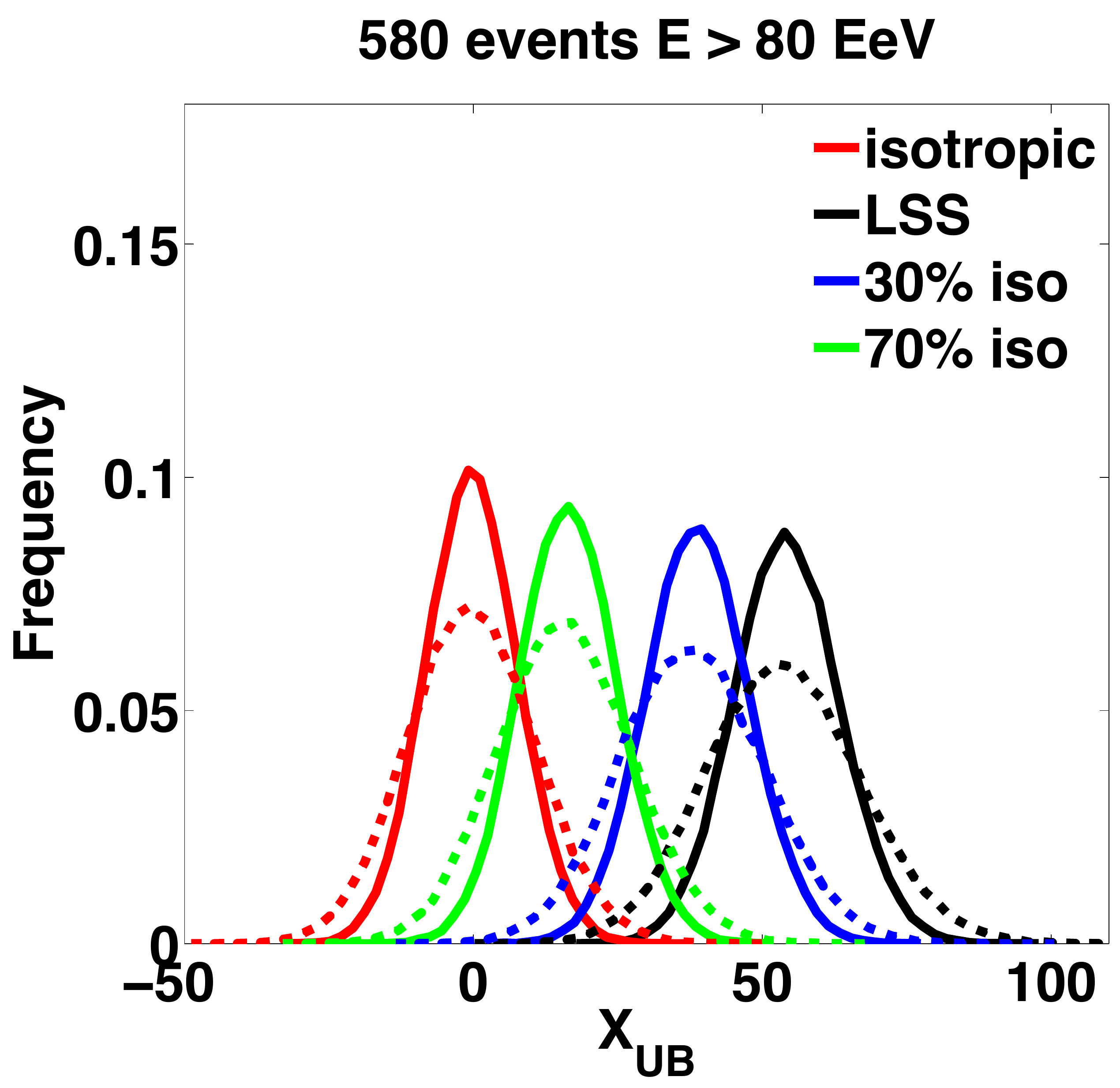}
\includegraphics[width=0.75 \figlength]{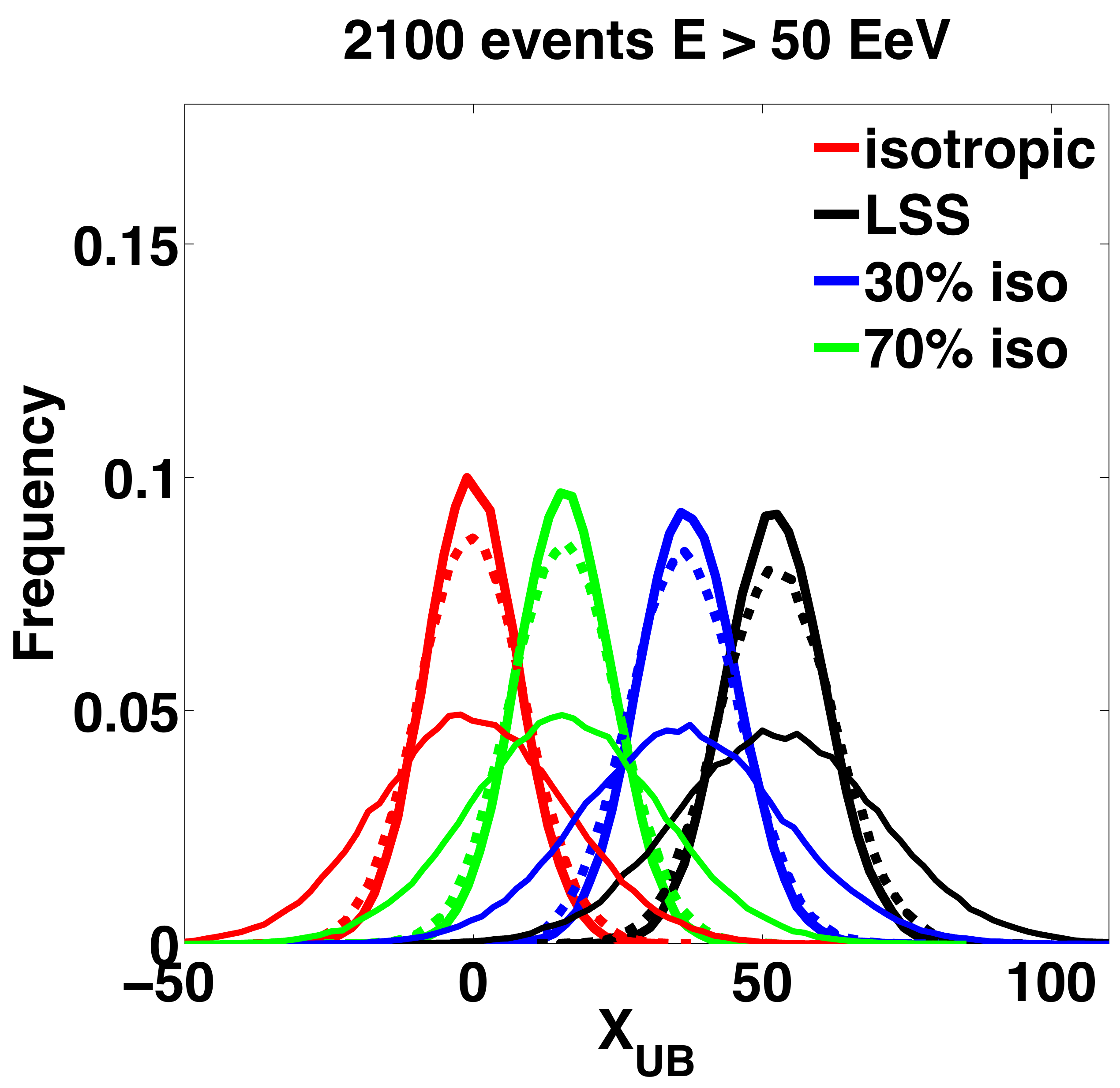}\\
\caption{The distribution of values of $X_{\rm UB}$ in $10^5$ realisations of UHECRs with energy $E \geq 100, 80, 50$~EeV from left to right. Red lines give the isotropic expectation, black lines the expectation from the unbiased model, blue (green) lines the expectation from the unbiased model, for a sample that is $30\%$ ($70\%$) isotropic. Thick solid lines correspond to a local UHECR source number density $\bar{n} = 10^{-2}~{\rm Mpc}^{-3}$,  dotted lines correspond to $\bar{n} = 10^{-3}~{\rm Mpc}^{-3}$ and thin solid lines to $\bar{n} = 10^{-4}~{\rm Mpc}^{-3}$ (the latter are only shown in the right column, see text for explanation). The top (bottom) row corresponds to the expected number of events for a total exposure $3 \times 10^{5}~{\rm km}^{2}~{\rm sr}~{\rm yr}$ following the Auger (TA) energy scale.}
\label{fig:X_UB_UNBIASED}
\end{figure}

Figure \ref{fig:X_UB_UNBIASED} shows the distribution of values of $X_{\rm UB}$ in $10^5$ Monte Carlo realisations of an isotropic model (black histograms), an unbiased model where $100\%$ of UHECRs are protons (red histograms), an unbiased model where $30\%$ of UHECRs arrive isotropically (blue histograms), and finally an unbiased model where $70\%$ of UHECRs arrive isotropically (green histograms).

It is clear that the expected anisotropy signal strongly depends on the number of events collected, and that it is significantly stronger if we consider all events collected with reconstructed energy $E \geq 50$~EeV than if we concentrate only on the smaller subset of the highest energy events. 

To quantify the anisotropy expected in the various scenarios for the UHECR source distribution considered, we follow the formalism developed in \cite{Waxman95} and quote the probability P($\bar{\rm M}_{1}| \rm M_{2}$) of ruling out a particular model of the source distribution $\rm M_{1}$ at a specified confidence level ($95\%,~99\%,~99.7\%$) assuming model $\rm M_{2}$ is true. We do so by considering the fraction of realisations of $\rm M_{1}$ in which $X_{\rm UB}$ took more extreme values than in $95\%,~99\%,~99.7\%$ of realisations of $\rm M_{2}$ respectively. 

\begin{figure}[!ht]
\centering
\includegraphics[width=1.2 \figlength]{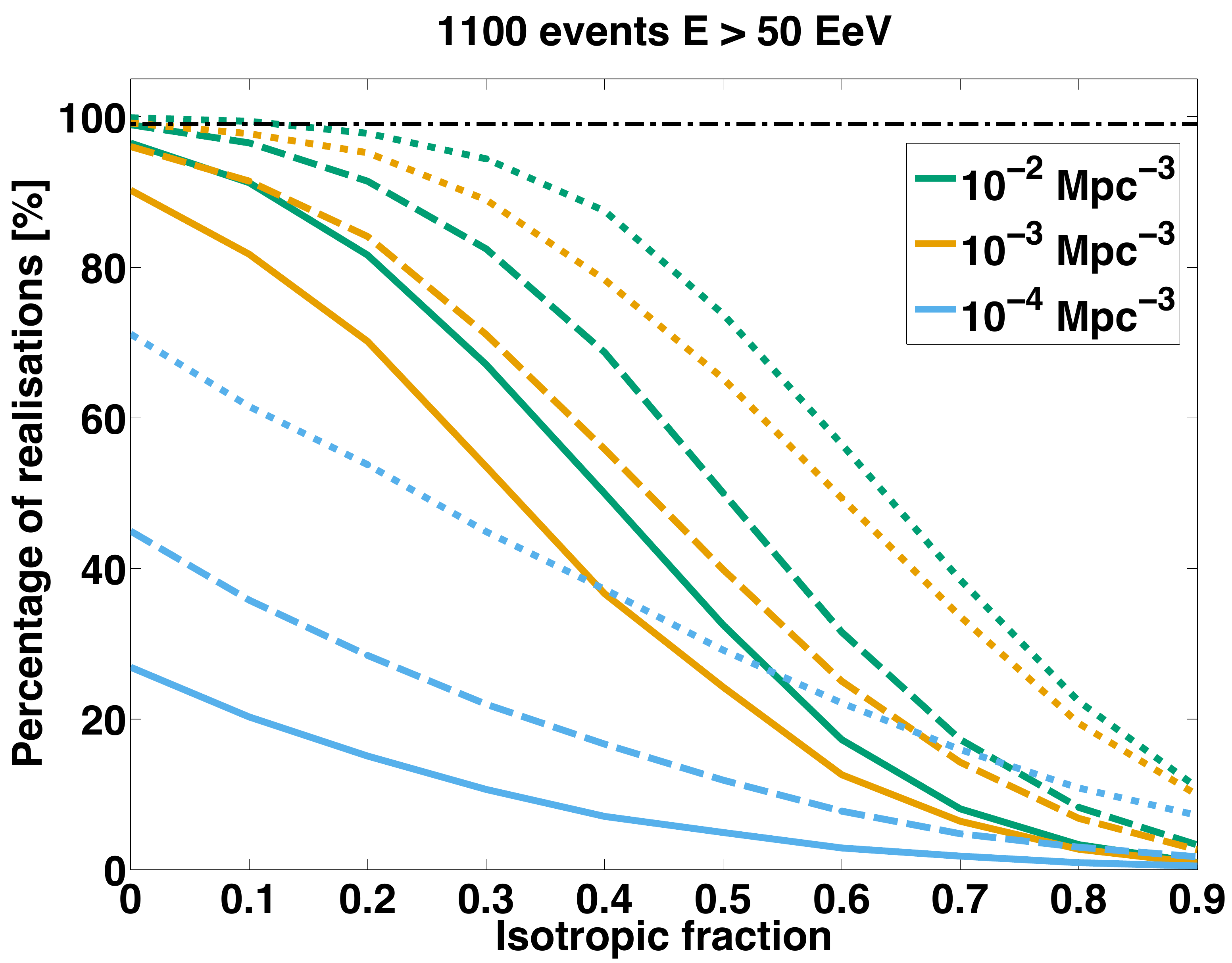}
\includegraphics[width=1.2 \figlength]{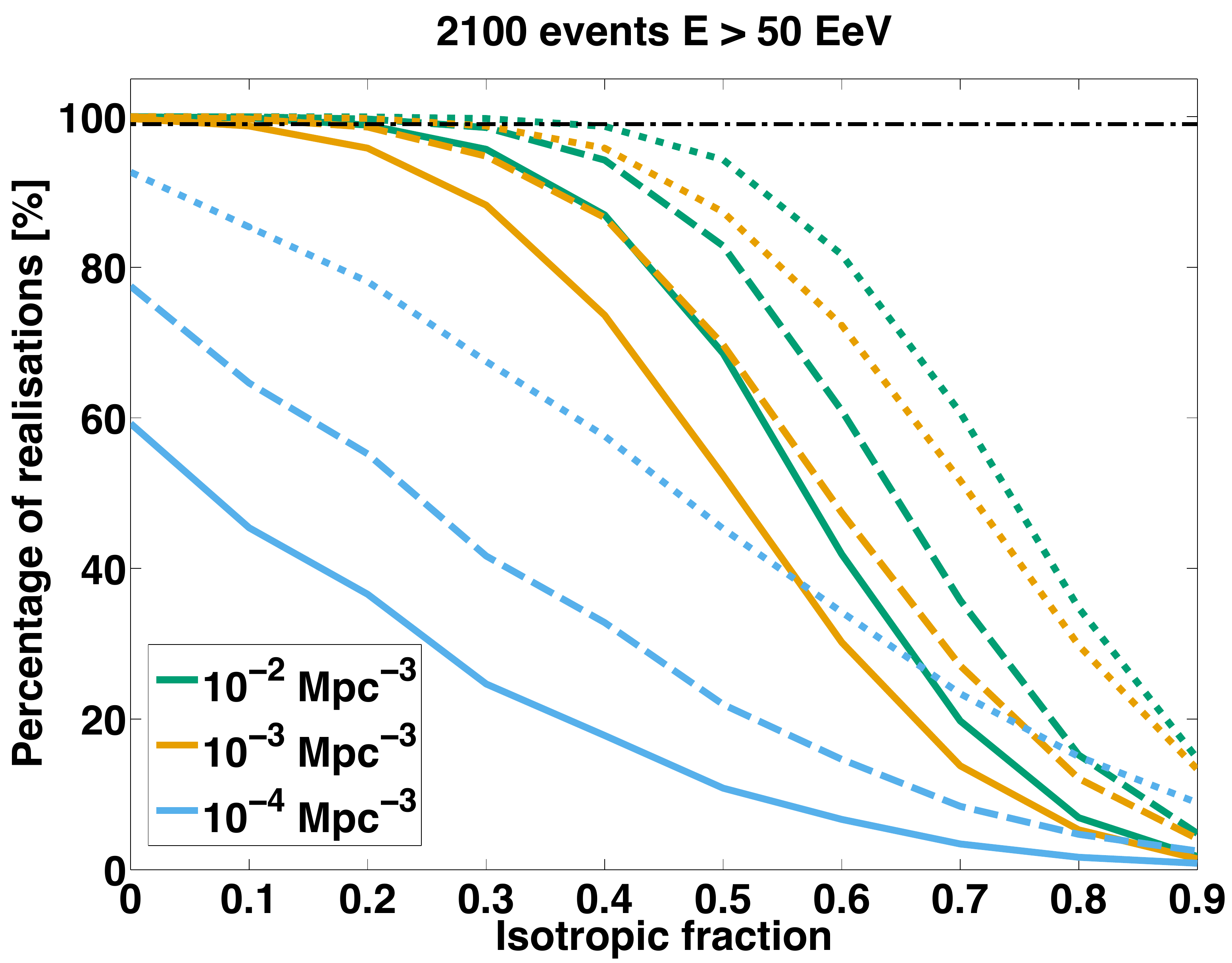}\\
\caption{The percentage of realisations of UHECRs from an {\it unbiased} source distribution, in which an anisotropy is expected with significance $ \geq 95\%$ (dotted lines), $ \geq 99\%$ (dashed lines), $ \geq 99.7\%$ (solid lines) as a function of the isotropic fraction of UHECRs present in the data, for the number of UHECRs expected with 5 years of JEM-EUSO, above $50$~EeV, assuming the Auger (TA) energy scale on the left (right) panel. The black dot-dashed horizontal line shows the $99\%$ CL. The UHECR source density is assumed to be $\bar{n} = 10^{-2}~{\rm Mpc}^{-3}$ (green lines), $\bar{n} = 10^{-3}~{\rm Mpc}^{-3}$ (orange lines), $\bar{n} = 10^{-4}~{\rm Mpc}^{-3}$ (light blue lines).}
\label{fig:SOURCE_DENSITY}
\end{figure}

Figure \ref{fig:SOURCE_DENSITY} shows the percentage of Monte Carlo realisations of an unbiased UHECR source model in which an anisotropy is expected with significance $ \geq 95\%$ (dotted lines), $ \geq 99\%$ (dashed lines), $ \geq 99.7\%$ (solid lines) as a function of the isotropic fraction of UHECRs present in the data, for the number of UHECRs expected with five years of JEM-EUSO above $50$~EeV, assuming the Auger (TA) energy scale on the left (right) panel. In other words, the quantity plotted on the $y$-axis of figure \ref{fig:SOURCE_DENSITY} is P$(\bar{\rm I}| \rm UB) \times 100$. The results plotted in figure \ref{fig:SOURCE_DENSITY} are based on the distribution of values of $X_{\rm UB}$ obtained in $10^5$ realisations of each of the UHECR source models considered. The black dot-dashed horizontal line shows the $99\%$ CL. 

Concentrating on the green lines, which correspond to a UHECR source distribution with local number density similar to that of bright galaxies, we see that an anisotropy at $>99\%$ ($>99.7\%$) CL is expected as long as the fraction of protons in the dataset is $\geq 70 \%$ ($\geq 80 \%$), when the number of UHECRs observed reaches 2100 above $50$~EeV. For lower source number densities a higher fraction of protons or a larger dataset are required for a statistically significant detection of anisotropy.

\subsection{Sensitivity to the UHECR source density}

The different linestyles in figure \ref{fig:X_UB_UNBIASED} show the distribution of values of $X_{\rm UB}$ obtained for different values of the UHECR number density. We observe, as expected (see discussion in \cite{kashti2008}), that the source number density does not affect the mean value of $X_{\rm UB}$ obtained for a specific model, but rather the width of the distribution of values of $X_{\rm UB}$ and therefore it does affect the significance with which a specific model for the UHECR source distribution can be ruled out. Further, we observe that a source number density similar to that of bright galaxies results in the strongest possible anisotropy signal and is thus the most favourable for the type of analysis we present. For an energy threshold of 80\,EeV or higher, we are unable to constrain source number densities lower than $10^{-3}~{\rm Mpc}^{-3}$ as the number of sources contributing to the expected UHECRs drops radically (the horizon over which the survival probability of $80$~EeV UHECRs drops to $1/e$ is $\sim60$~Mpc making the number contributing UHECR sources $(4/3)\pi(60^3)\bar{n} \sim 200$, for $\bar{n} = 10^{-4}~{\rm Mpc}^{-3}$). As a result, the distribution of $X_{\rm UB}$ becomes highly non-Gaussian, not allowing us to derive unambiguous confidence intervals. For a study of this low source number density regime, see the work of \cite{Blaksley13}. 

Figure \ref{fig:SOURCE_DENSITY} shows the significance with which it will be possible to rule out anisotropy once the number of observed events exceeds 1000 beyond $E > 50$~EeV, for different mean values of the UHECR source density, $\bar{n}$, assuming the UHECR source distribution follows that of the unbiased model. We observe that a lower value of $\bar{n}$ degrades the expected anisotropy signal as expected based on the results of figure \ref{fig:X_UB_UNBIASED}. Inspection of the light blue lines in figure \ref{fig:SOURCE_DENSITY} shows that if $\bar{n} \leq 10^{-4}~{\rm Mpc}^{-3}$, a correlation with the matter distribution will not be detectable with high significance, unless the composition is proton dominated up to the highest energies and either (a) the number of observed events significantly exceeds 2000 or (b) the distribution of UHECR sources is strongly clustered, as in the threhold bias model (see \ref{subsec:bias} for details). A detection of anisotropy is on the other hand expected with 2100 events above $50$~EeV, even for the unbiased source model, if $\bar{n} \gtrsim 10^{-3}~{\rm Mpc}^{-3}$, as long as protons constitute $\geq 90\%$ of UHECRs, by number, in this energy range.

\subsection{Sensitivity to the bias of the source distribution}
\label{subsec:bias}

\begin{figure}[!ht]
\centering
\includegraphics[width=0.75 \figlength]{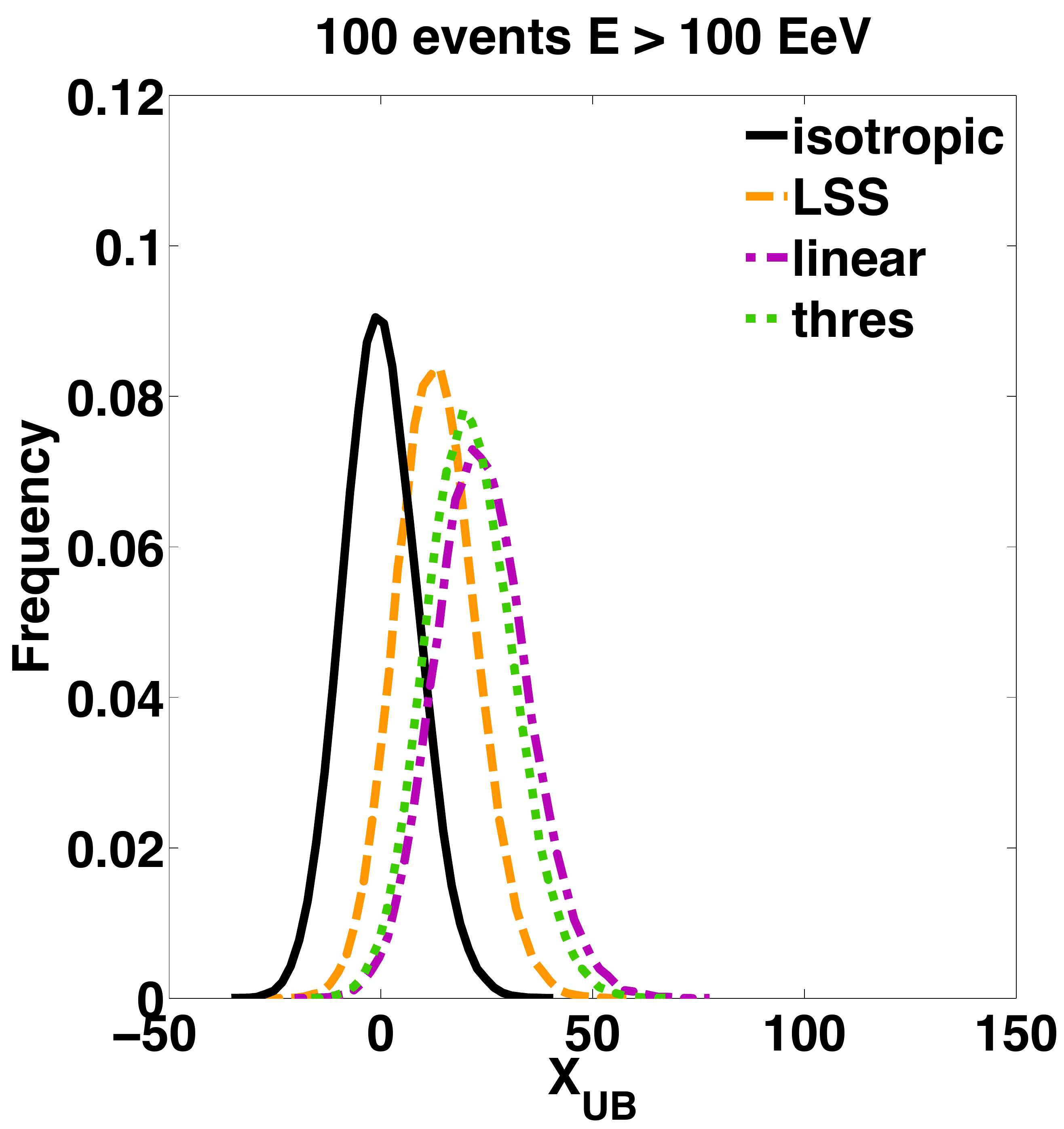}
\includegraphics[width=0.75 \figlength]{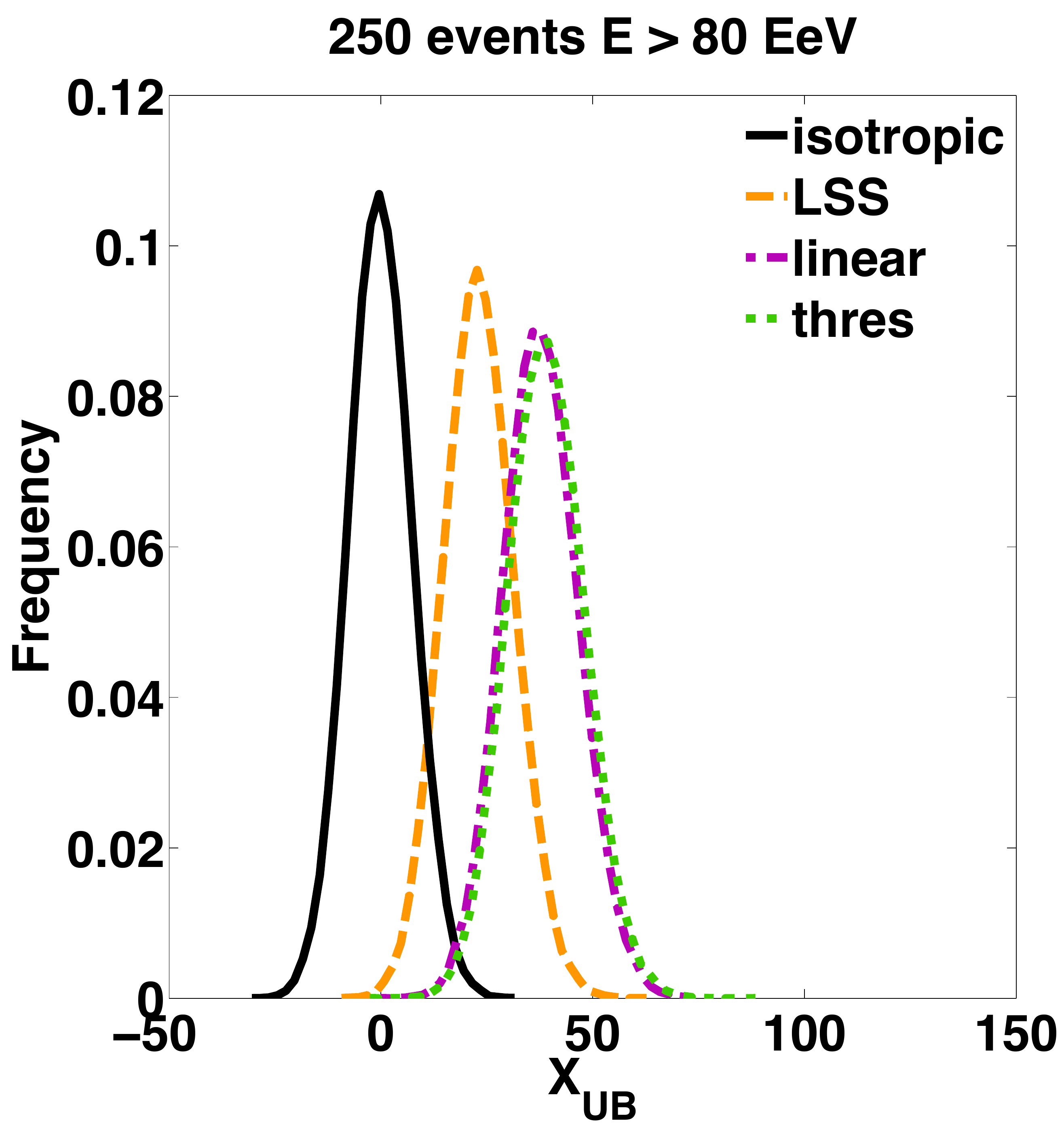}
\includegraphics[width=0.75 \figlength]{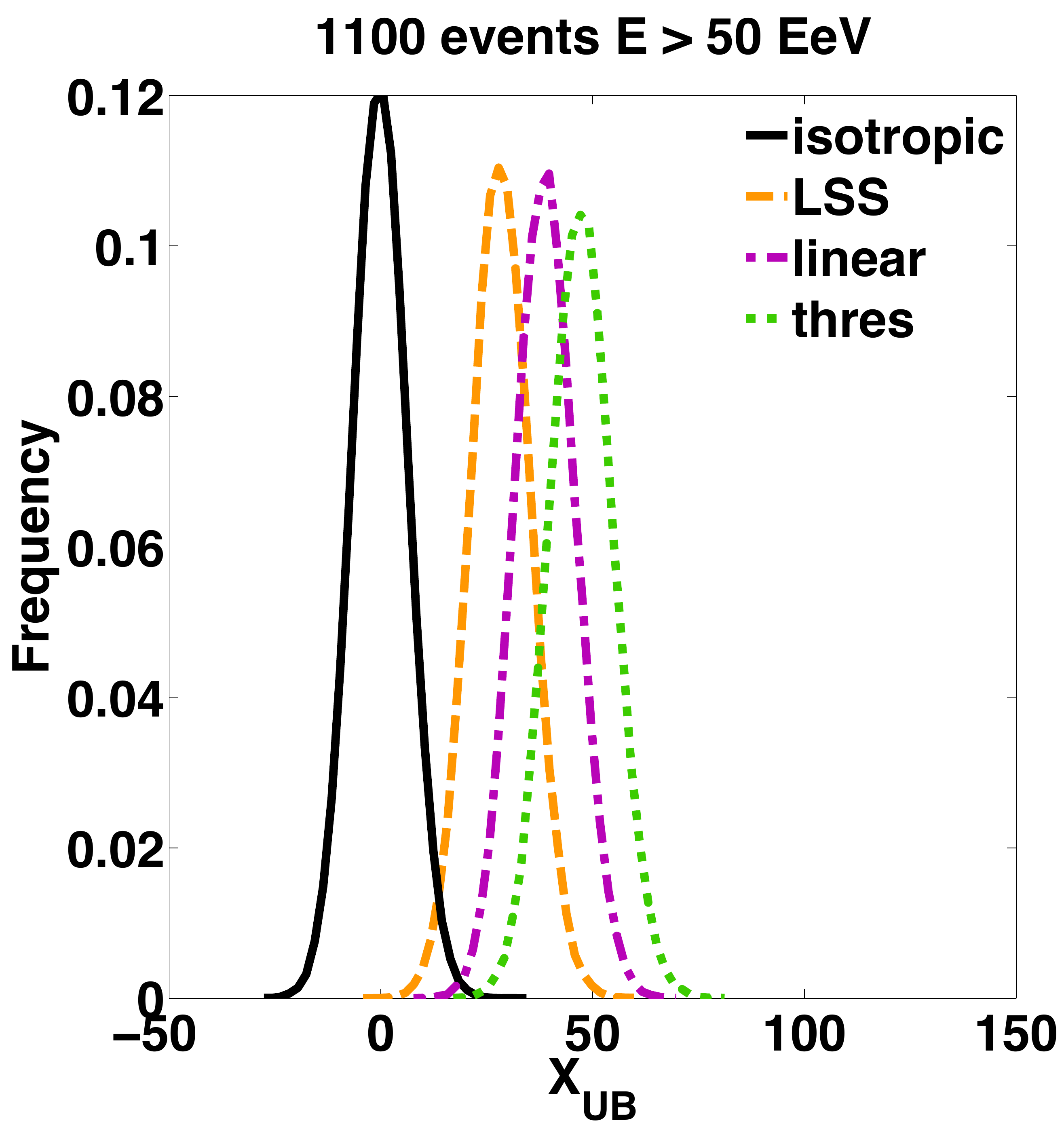}\\
\includegraphics[width=0.75 \figlength]{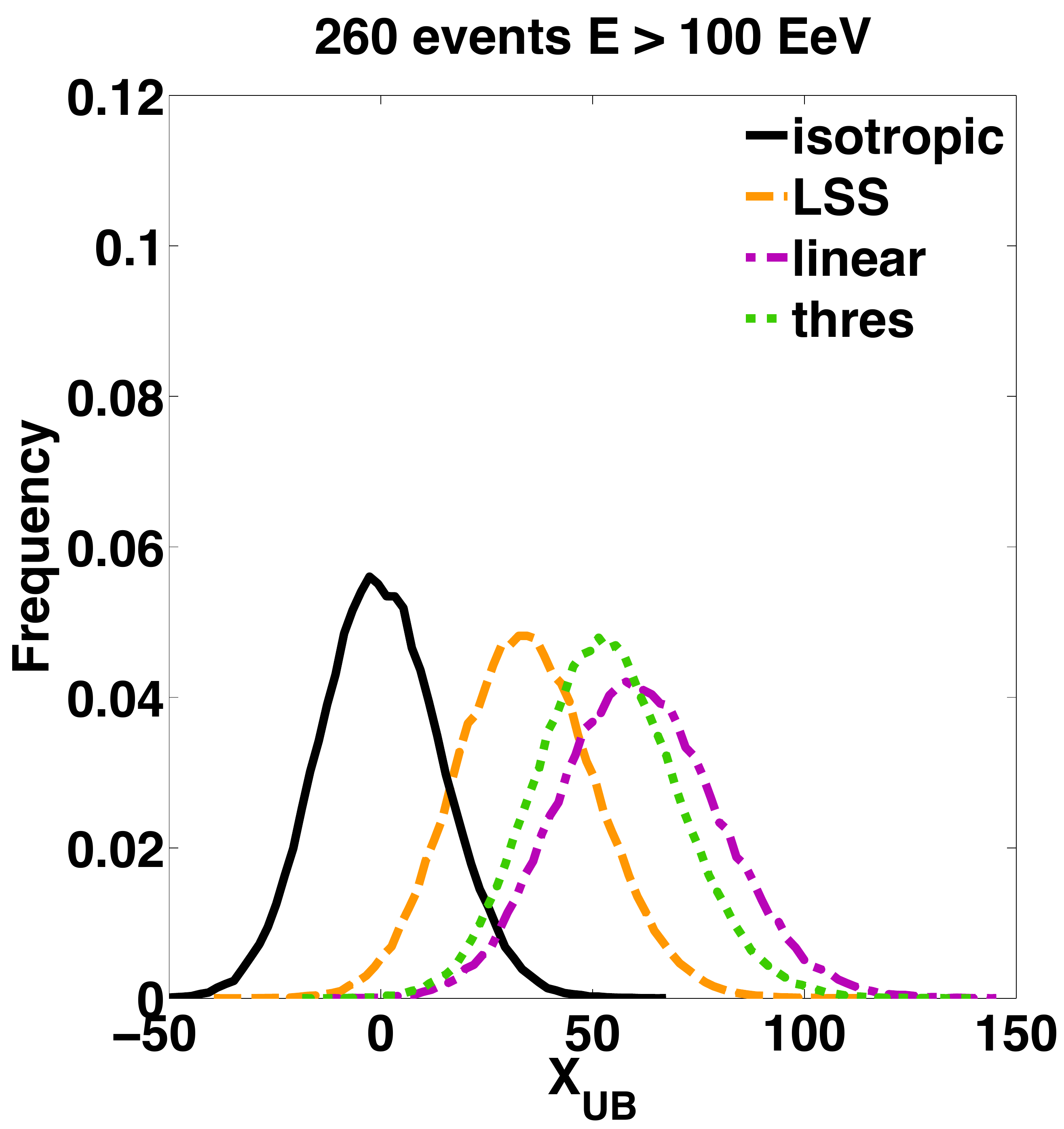}
\includegraphics[width=0.75 \figlength]{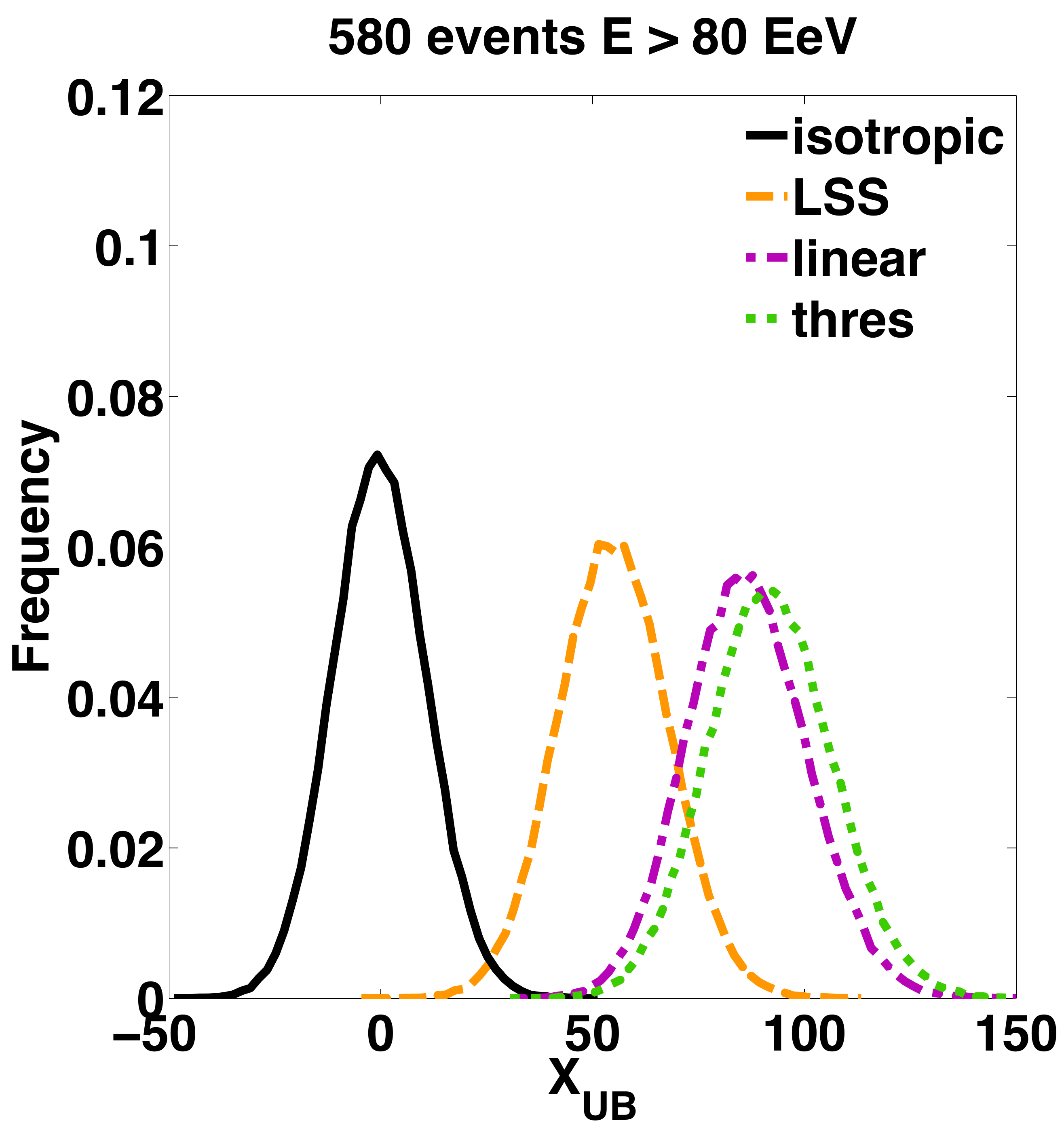}
\includegraphics[width=0.75 \figlength]{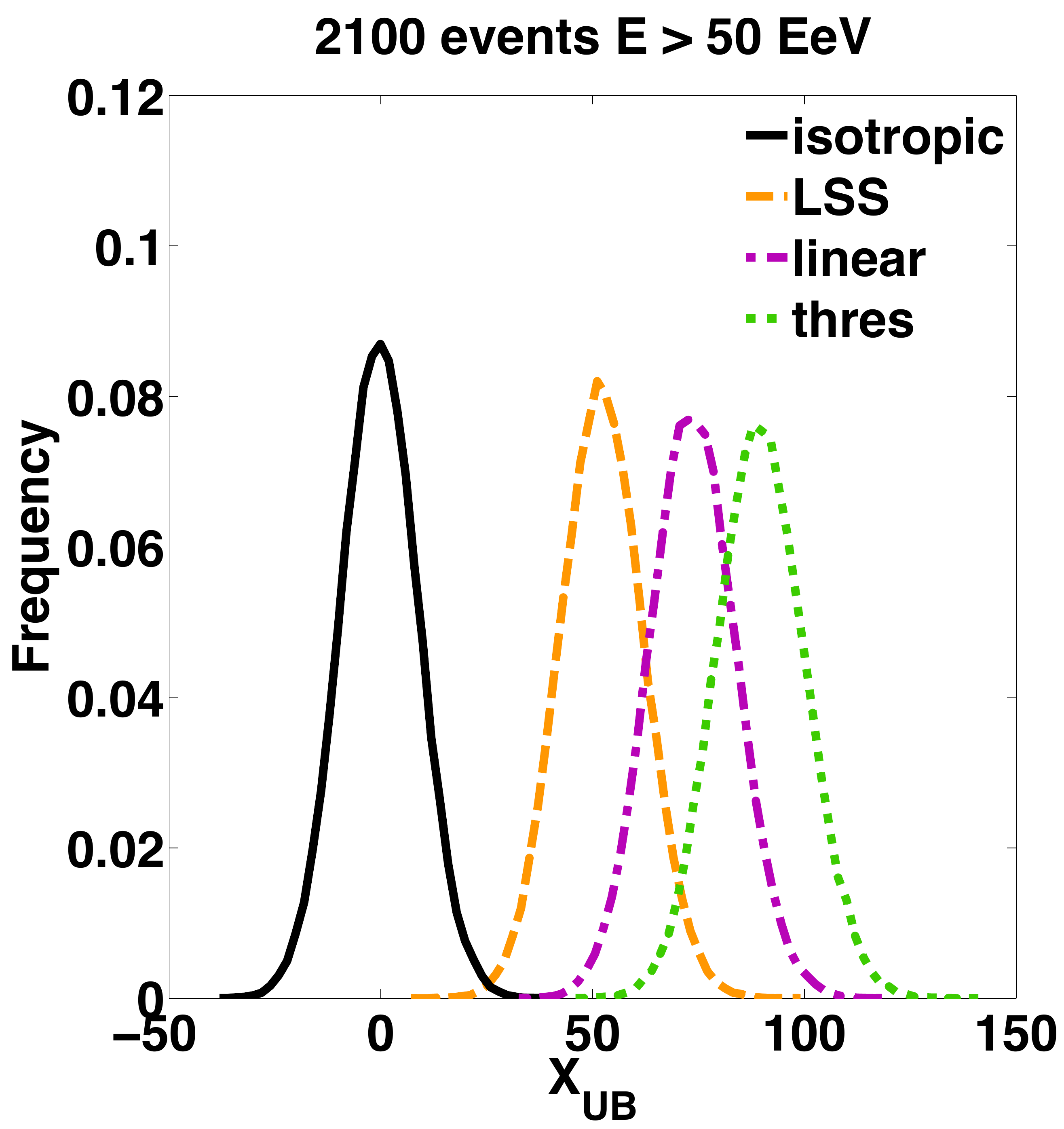}\\

\caption{The distribution of values of $X_{\rm UB}$ in $10^5$ realisations of UHECRs with energy $E \geq 100, 80, 50$~EeV from left to right in the different bias models. Black histograms give the isotropic expectation, orange histograms the expectation from the unbiased model (marked LSS in the legend), purple histograms the expectation from the linear bias model and green histograms the expectation from the threshold bias model. The expected number of events shown in the top (bottom) row corresponds to the expected number of detected events with 5 years of JEM-EUSO following the Auger (TA) energy scale. The UHECR source number density has been assumed to be $\bar{n} = 10^{-3}~{\rm Mpc}^{-3}$.}
\label{fig:X_UB_ALL}
\end{figure}

\begin{figure}[!ht]
\centering
\includegraphics[width=0.75 \figlength]{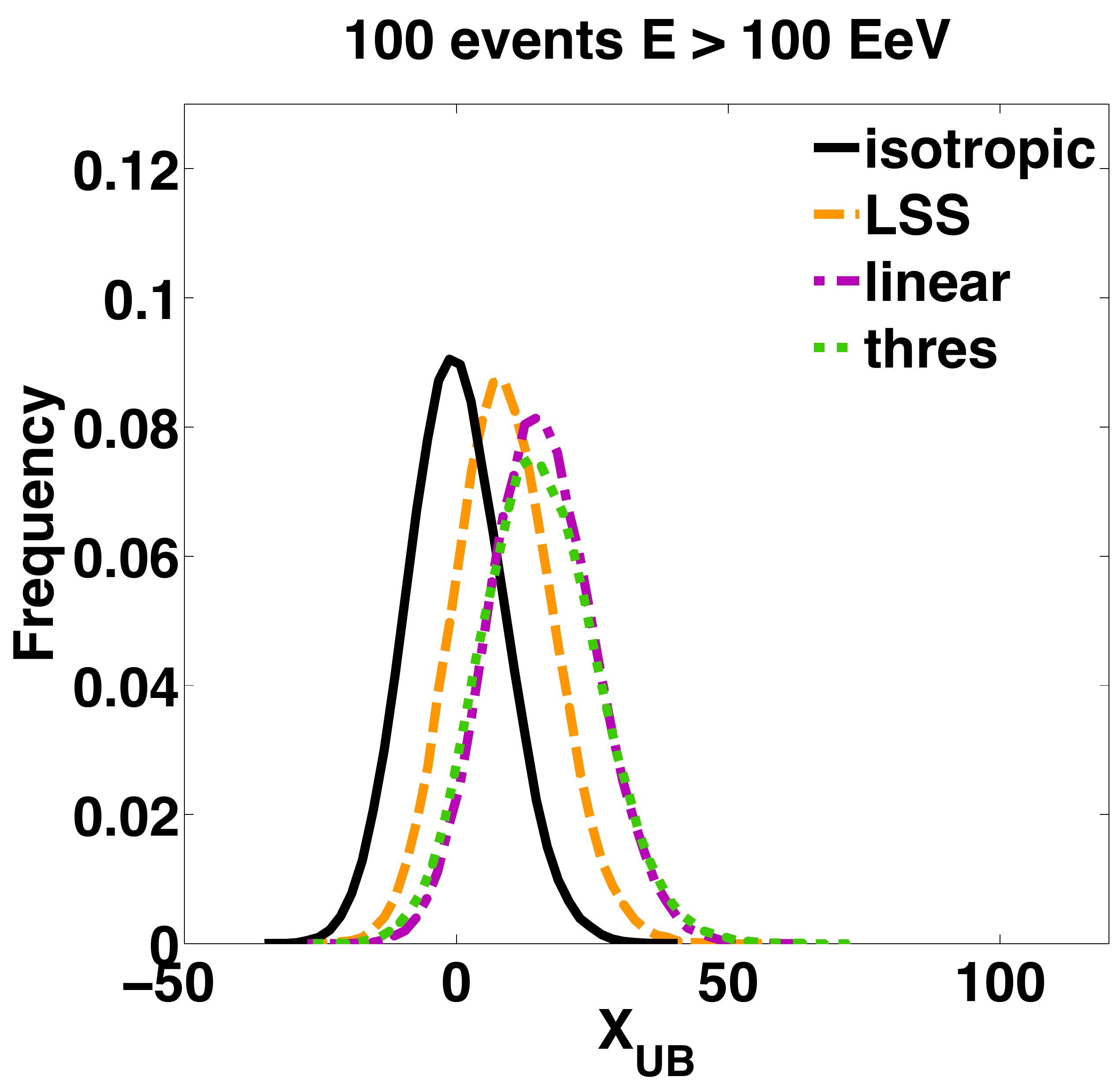}
\includegraphics[width=0.75 \figlength]{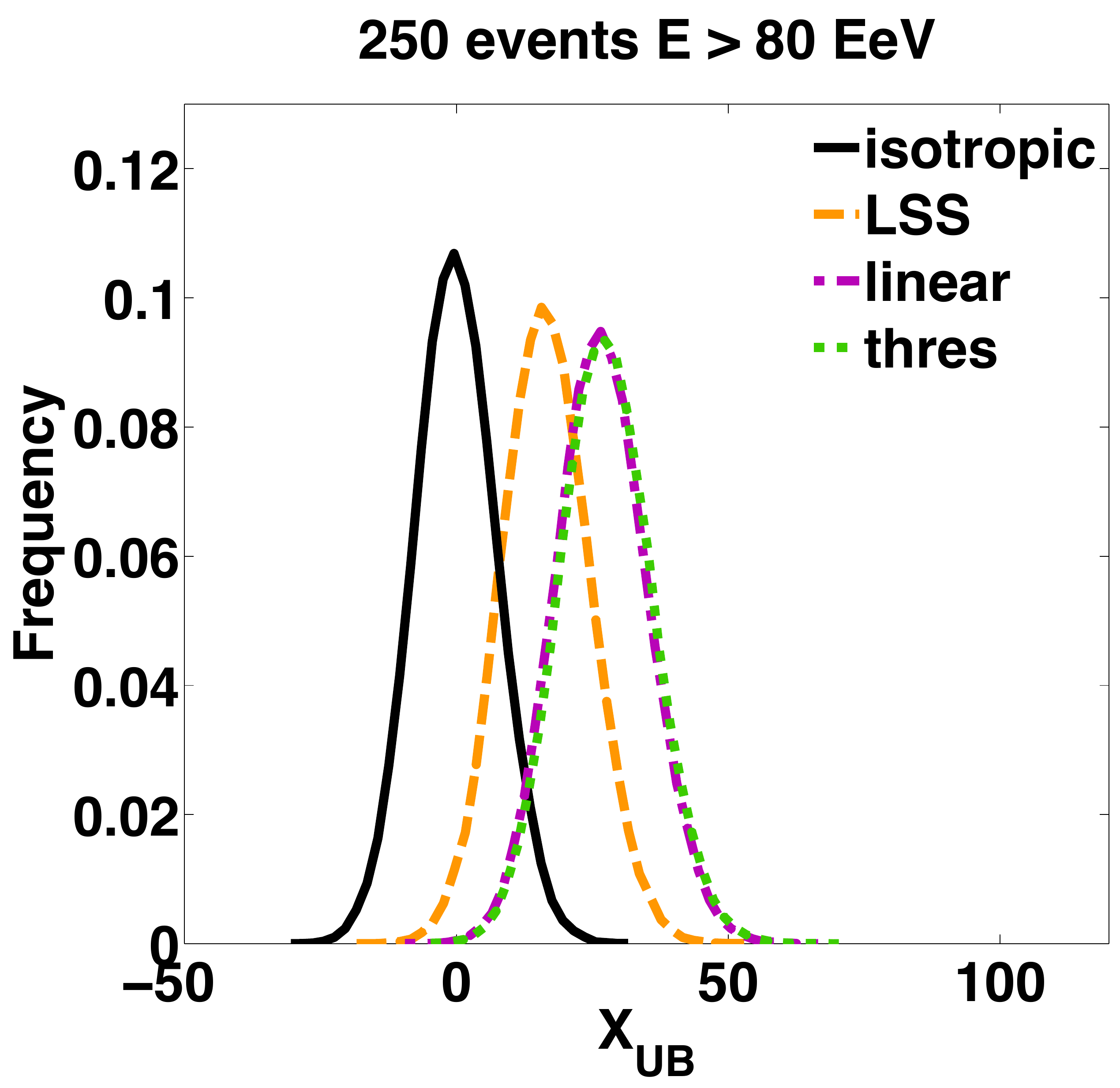}
\includegraphics[width=0.75 \figlength]{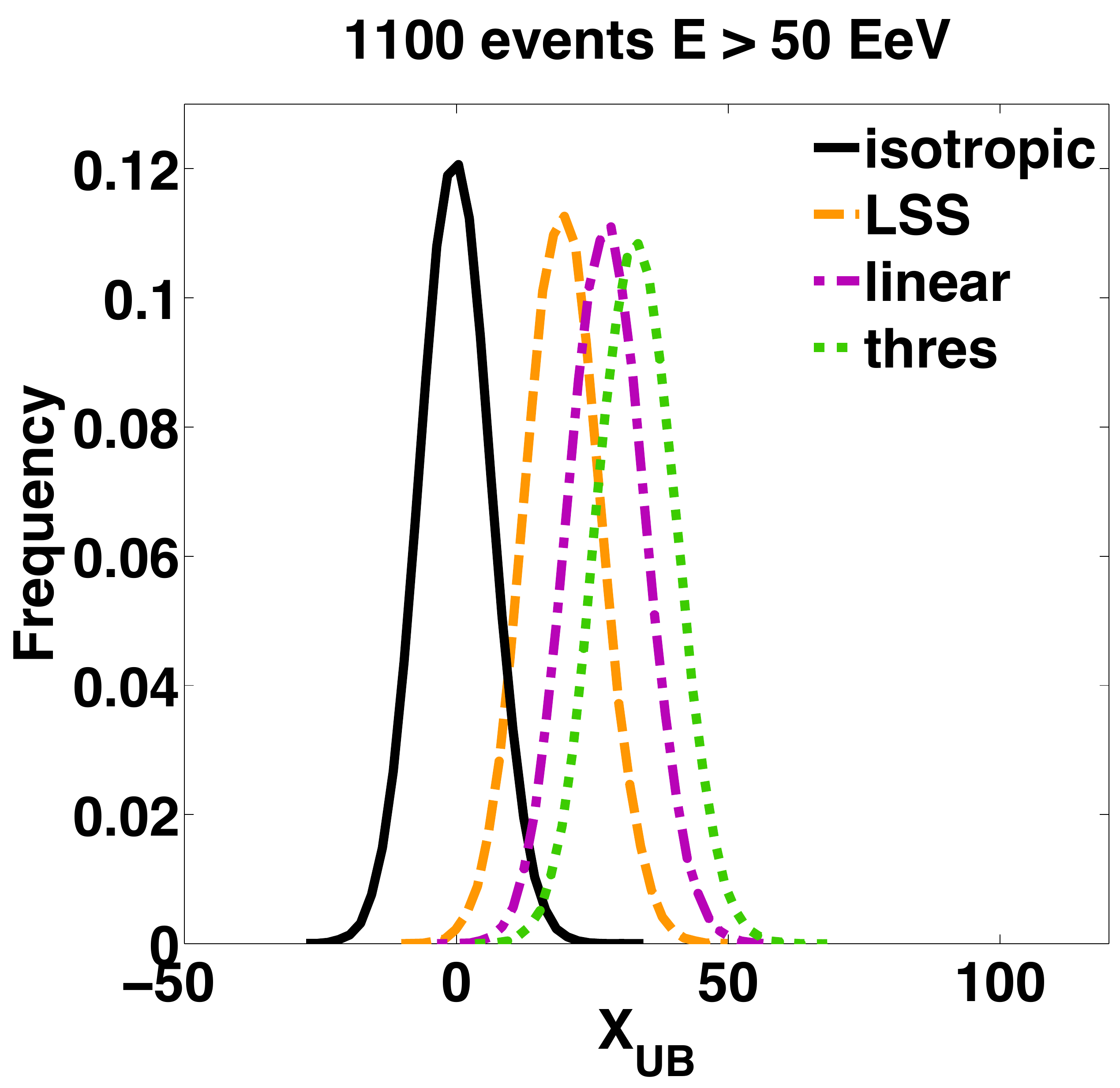}\\
\includegraphics[width=0.75 \figlength]{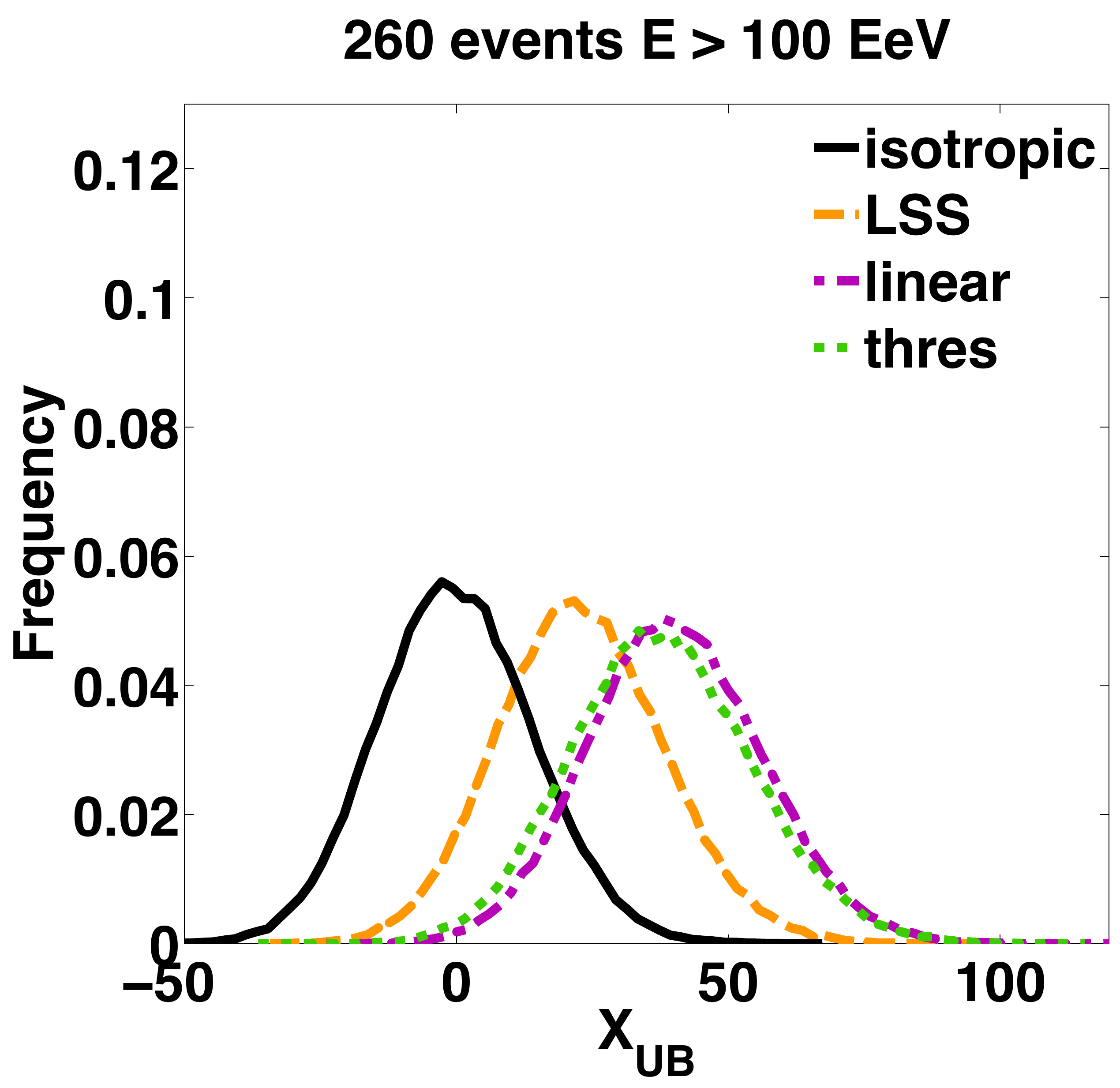}
\includegraphics[width=0.75 \figlength]{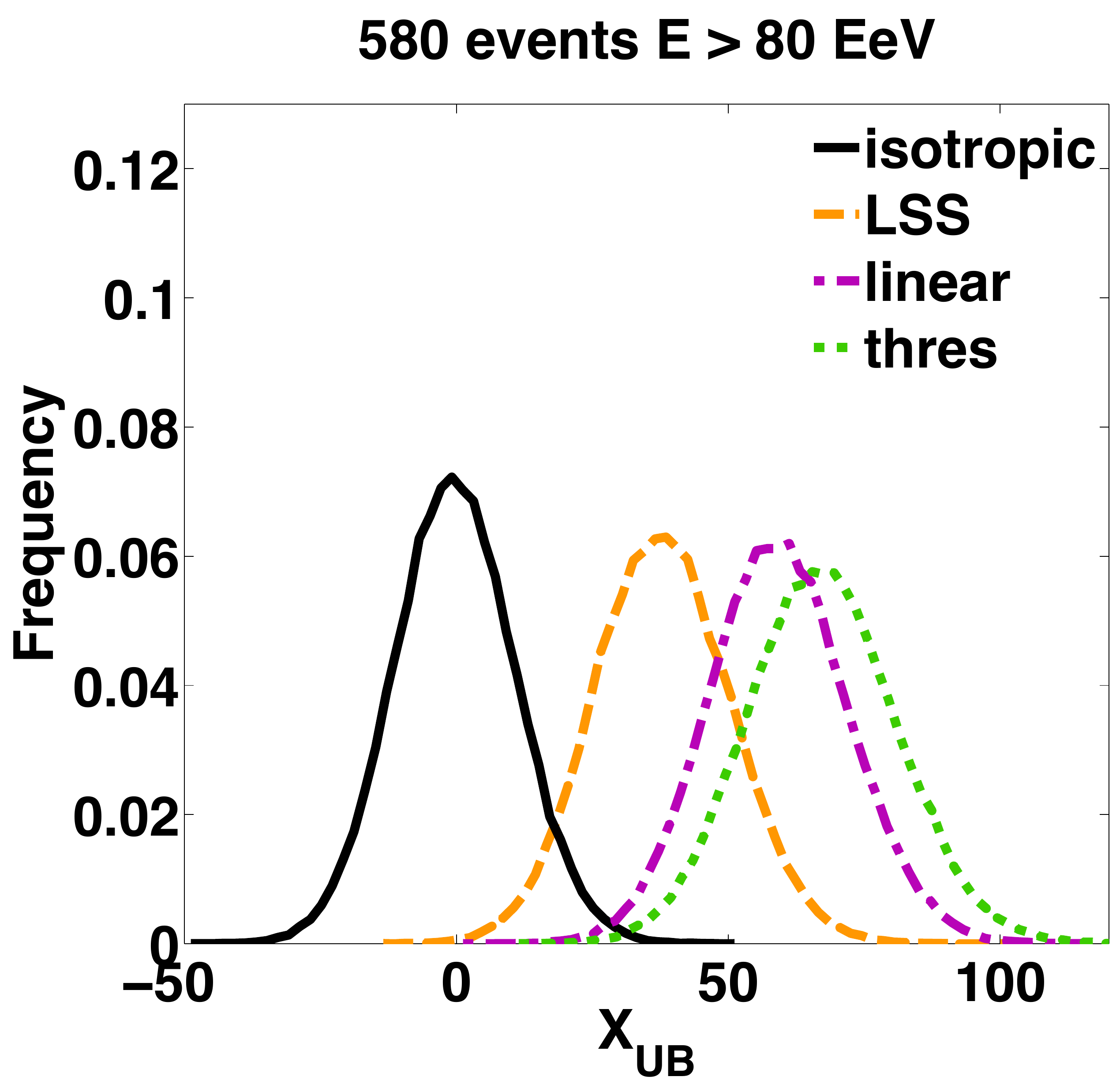}
\includegraphics[width=0.75 \figlength]{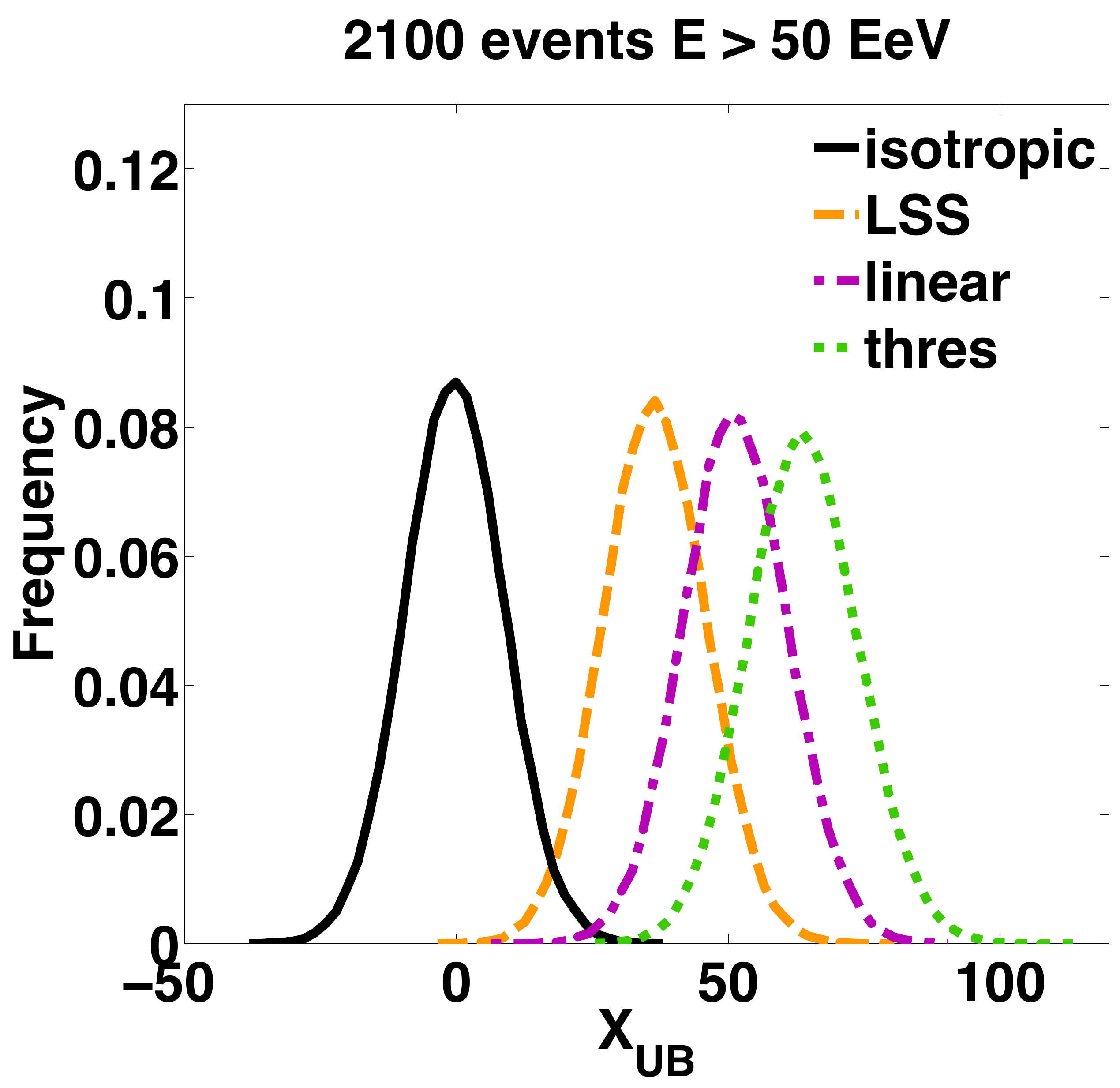}
\caption{Same as figure \ref{fig:X_UB_ALL} but assuming $30\%$ of detected UHECRs arrive isotropically.}
\label{fig:X_UB_ALL_ISO_30}
\end{figure}

\begin{figure}
\centering
\includegraphics[width=1.05 \figlength]{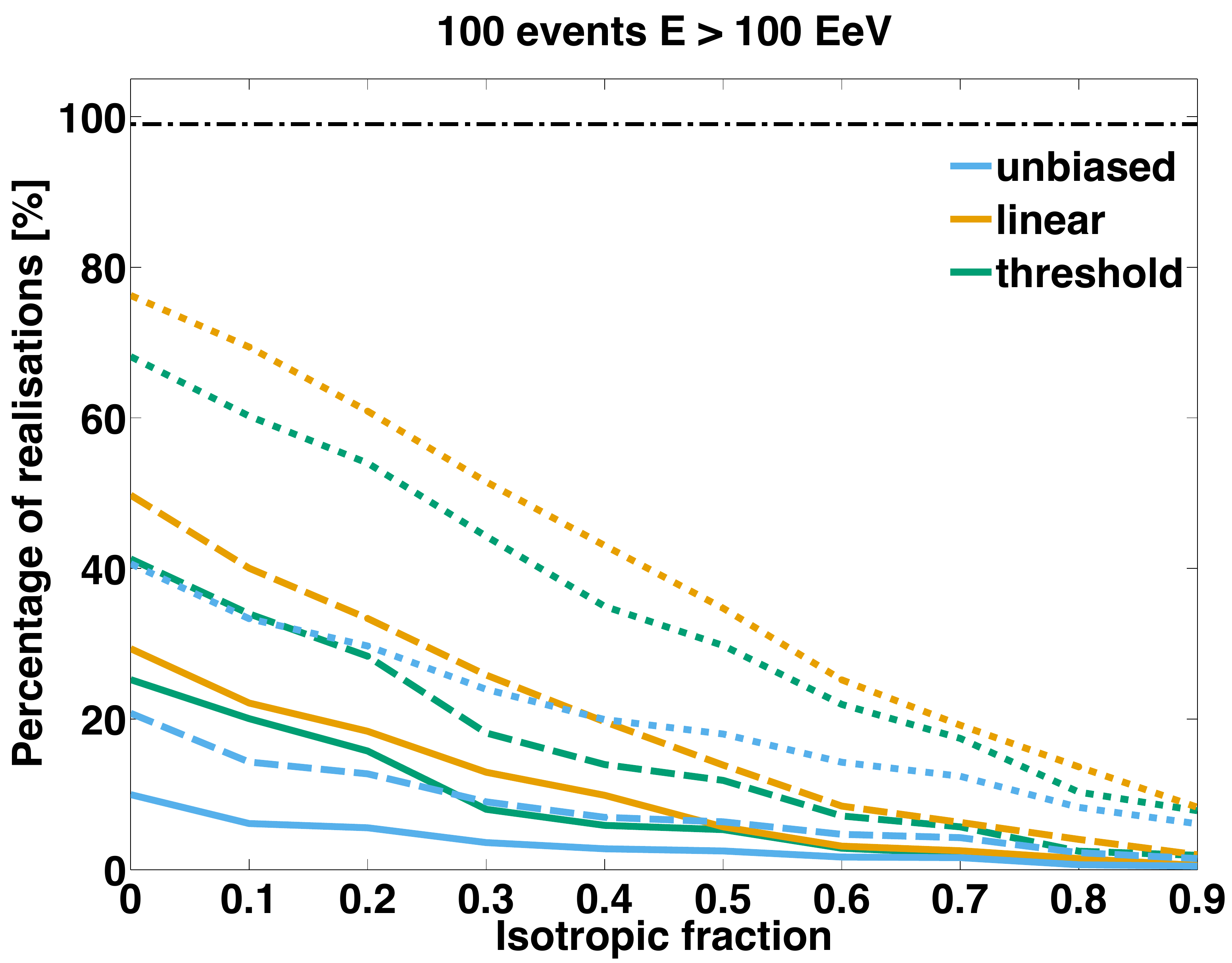}
\includegraphics[width=1.05 \figlength]{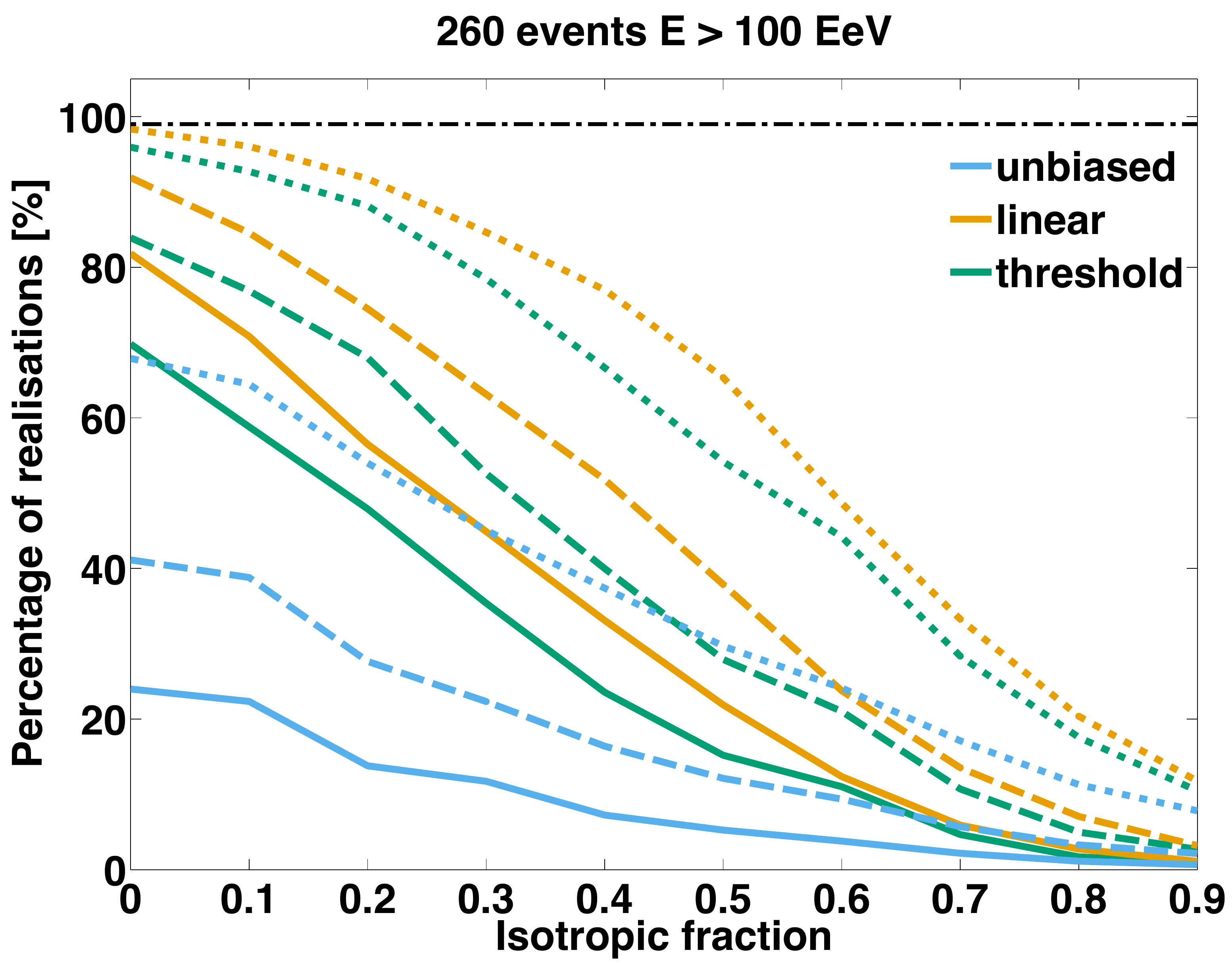}\\
\includegraphics[width=1.05 \figlength]{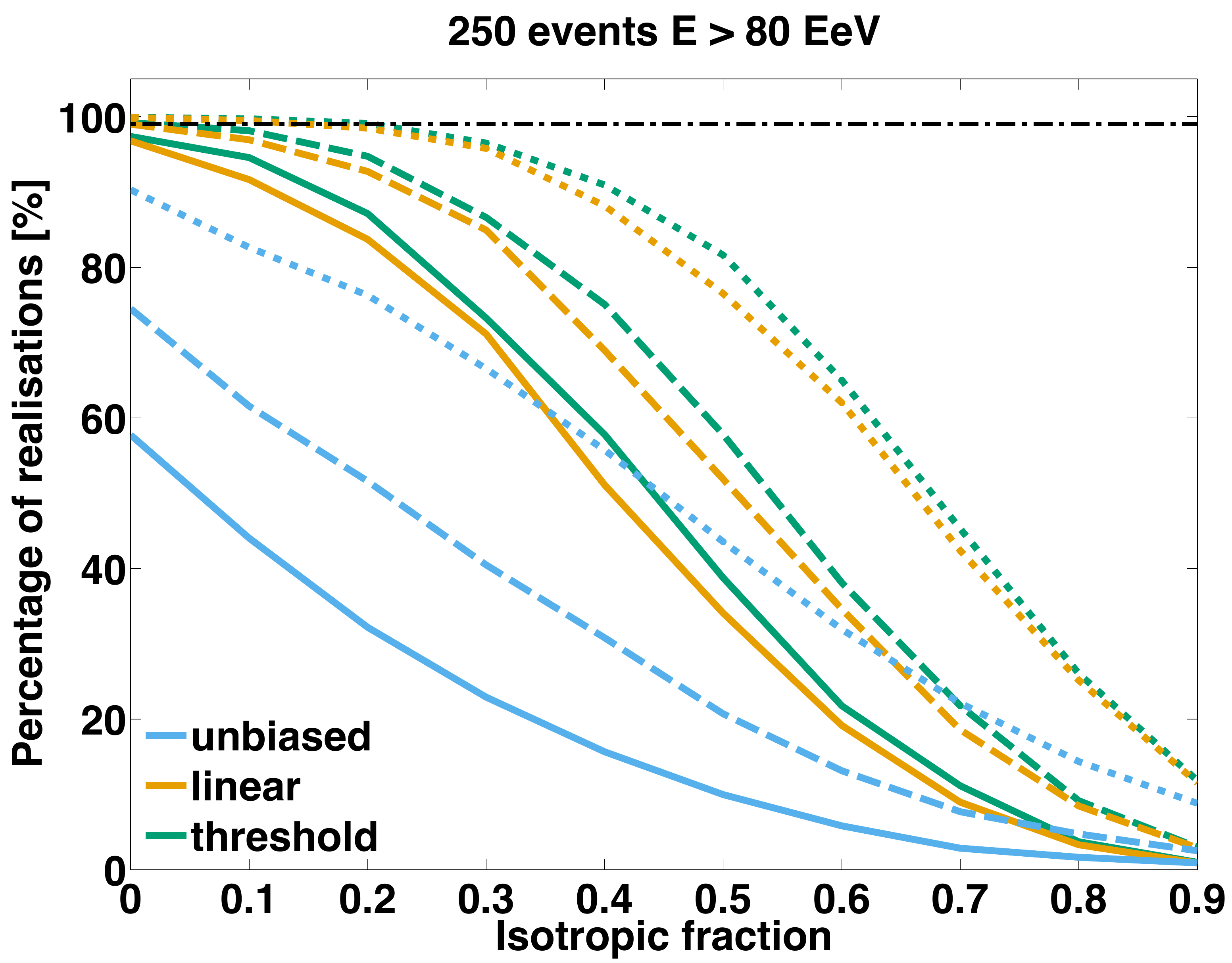}
\includegraphics[width=1.05 \figlength]{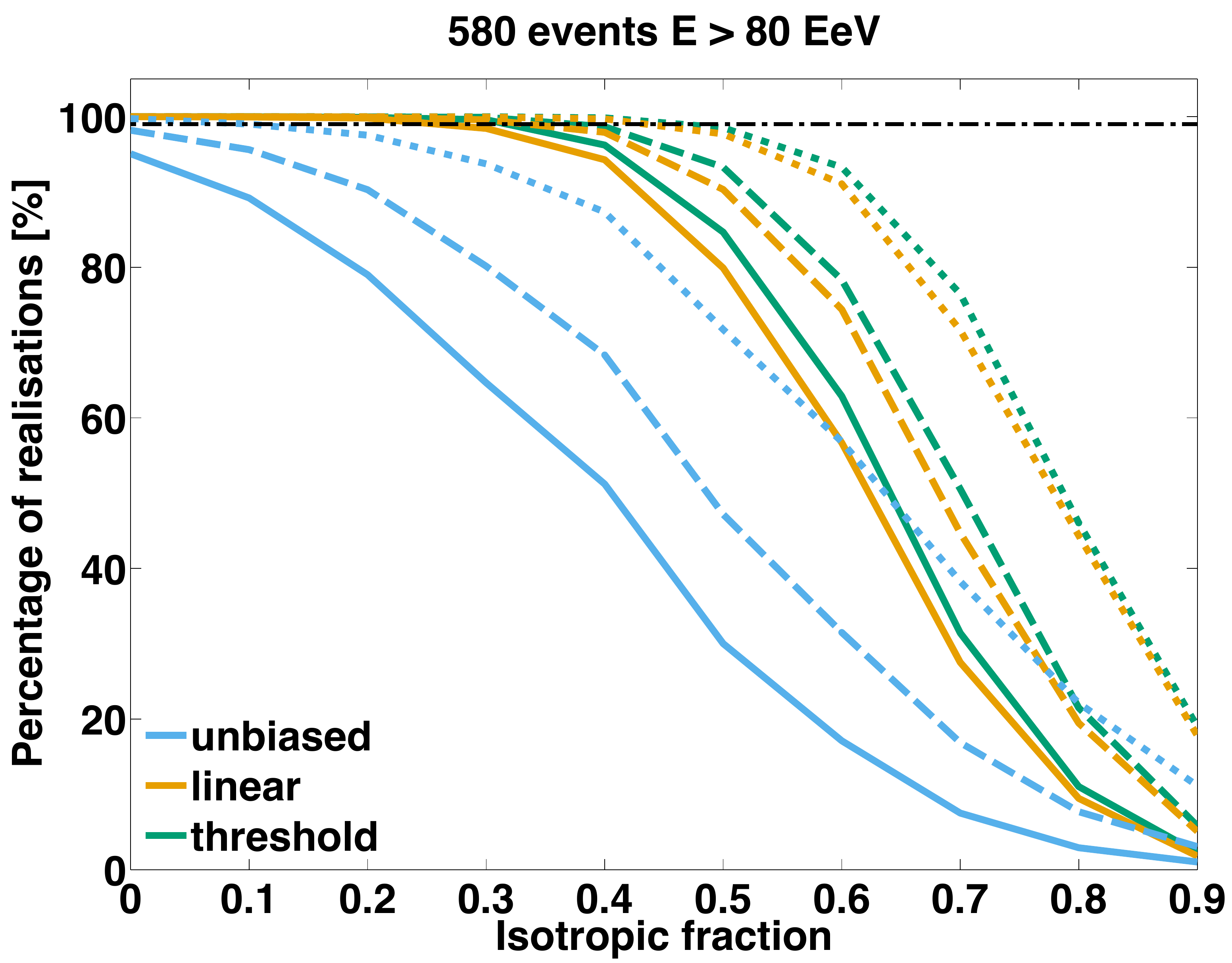}\\
\includegraphics[width=1.05 \figlength]{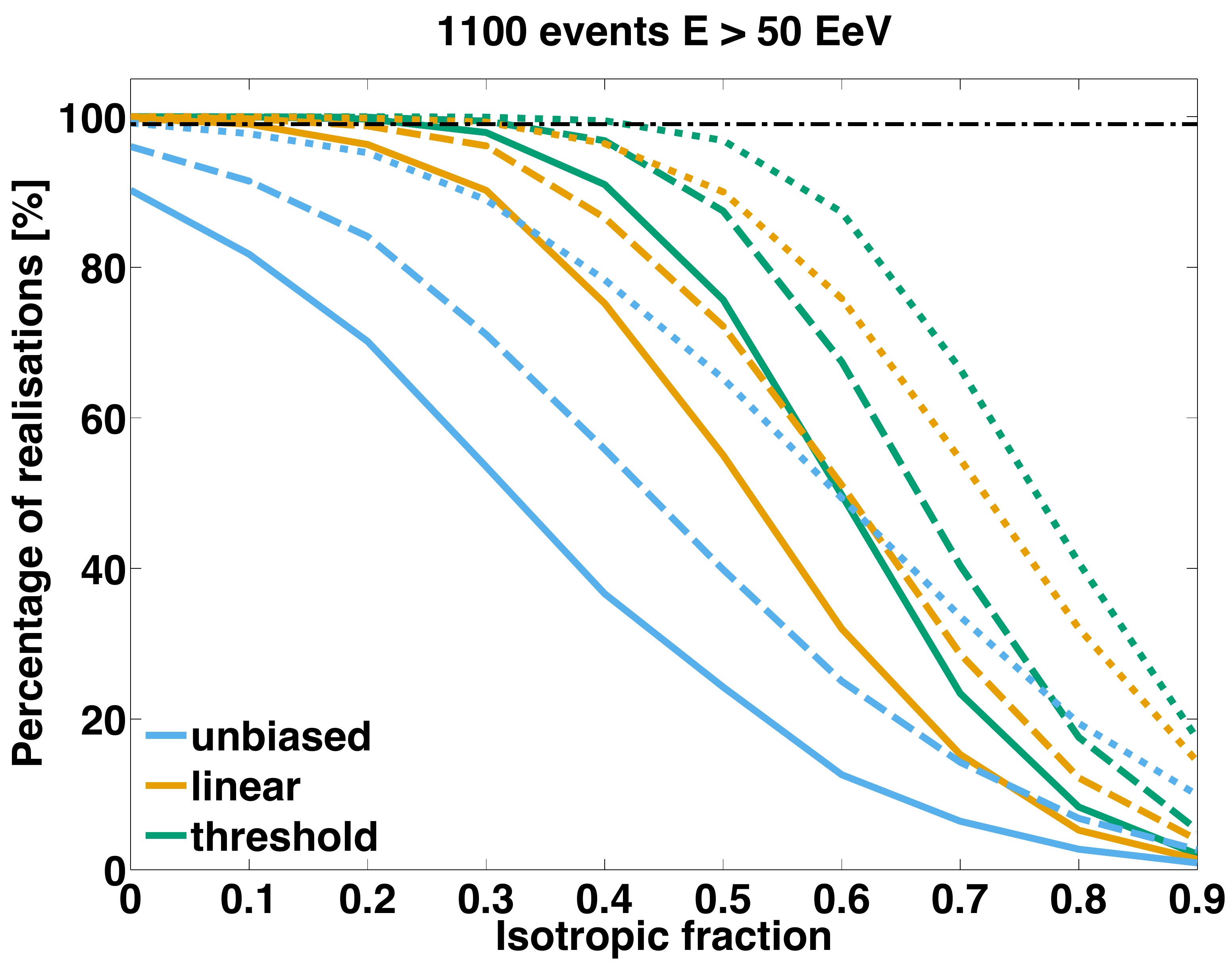}
\includegraphics[width=1.05 \figlength]{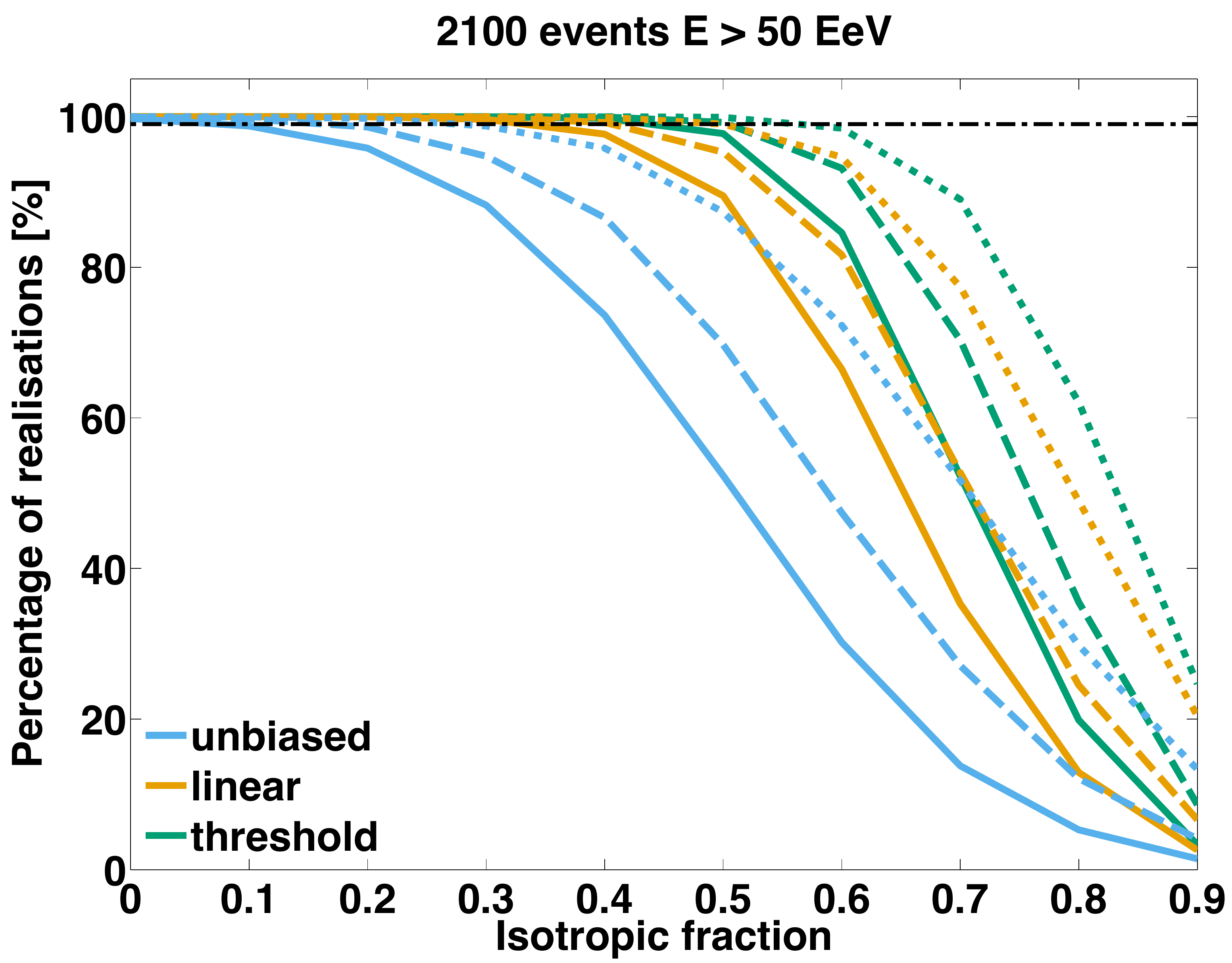}\\
\caption{The percentage of realisations of one of the models assumed for the UHECR source distribution (unbiased-light blue, linear bias-orange, threshold bias-green) in which an anisotropy is expected with significance $ \geq 95\%$ (dotted lines), $ \geq 99\%$ (dashed lines), $ \geq 99.7\%$ (solid lines) as a function of the isotropic fraction of UHECRs present in the data. The black, dot-dashed horizontal line shows the $99\%$ CL. The UHECR source density is assumed to be $\bar{n} = 10^{-3}~{\rm Mpc}^{-3}$. The top, bottom and middle rows correspond to predictions based on the number of evens expected with energy beyond $E \geq 100, 80, 50$~EeV respectively.}
\label{fig:X_UB_AS_FN_ISO}
\end{figure}

We now look at the sensitivity of a future UHECR detector to the anisotropy signal expected in the different clustering models that we have assumed for the cosmic ray sources. With the same setup as previously, we calculate the distribution of values of $X_{\rm UB}$ in the different possible models of the source distribution and plot these in figures \ref{fig:X_UB_ALL}-\ref{fig:X_UB_ALL_ISO_30}. As in our earlier results, we observe that the much larger number of lower energy events is crucial for distinguishing between the different bias models. In the top, right panel of figure \ref{fig:X_UB_ALL}, we see that already with 1100 events there is significant distinction between the isotropic, unbiased and linear/threshold models. The situation only improves, if we assume the TA energy scale. 

Further, inspection of figures \ref{fig:X_UB_ALL} and \ref{fig:X_UB_ALL_ISO_30} reveals that the trend for the mean of the distribution of $X_{\rm UB}$ changes in a similar way as we change the bias model and as we change the isotropic fraction in the data. In other words, with future data we will be able to constrain the 2D parameter space defined by the combination of the composition and the bias model with the statistic $X_{\rm UB}$ alone. Inspection of the histograms in figures \ref{fig:X_UB_ALL}-\ref{fig:X_UB_ALL_ISO_30} leads us to a further, slightly disappointing conclusion that the mean value of the cross-correlation will have degeneracies as a result of the effect of the unknown composition/deflections of UHECRs and that of the unknown bias of the UHECR sources with respect to the galaxy distribution. Only knowledge of the composition of the UHECR sample studied, will allow to break such a degeneracy. 

Despite this drawback, our main concern is whether a significant anisotropy should be expected or not. In figure \ref{fig:X_UB_AS_FN_ISO}, we quantify the expected anisotropy signal in the models presented in figures \ref{fig:X_UB_ALL}-\ref{fig:X_UB_ALL_ISO_30}, as a function of the isotropic fraction of events present in the data. In the bottom row, we see that when the number of detected events reaches 2100 an anisotropy at $\geq 99.7\%$ CL is expected in all of the bias models, if the fraction of protons is $\geq 90\%$. Inspection of the green lines confirms that a very strong anisotropy signal is expected in the threshold bias model, even for a large fraction of isotropised events: namely an anisotropy at $>99.7\%$ CL is expected, even if the fraction of protons is as low as $60\%$ ($80\%$), when the number of observed events reaches 2100 (1100). A more moderate anisotropy signal is expected in the linear bias model, where the fraction of protons should be $\geq 70\%$ for 2100 observed events, in order to be able to rule out isotropy at the $>99.7\%$ CL.

The confidence intervals quoted here can be considered a firm lower limit to the expected anisotropy signal for the models considered, as our assumption that heavier nuclei than protons arrive isotropically certainly results in a conservative estimate of the expected anisotropy. We also note that our conclusions for the full proton composition are more conservative than those of \cite{kashti2008} for a specified number of events, as a result of having taken more conservative values for the energy and pointing resolution of the experiment, given the prospect of it being an instrument in space. 

\subsection{Results for sub-sample of light elements}
\label{subsec:bias}

\begin{figure}
\centering
\includegraphics[width=1.075 \figlength]{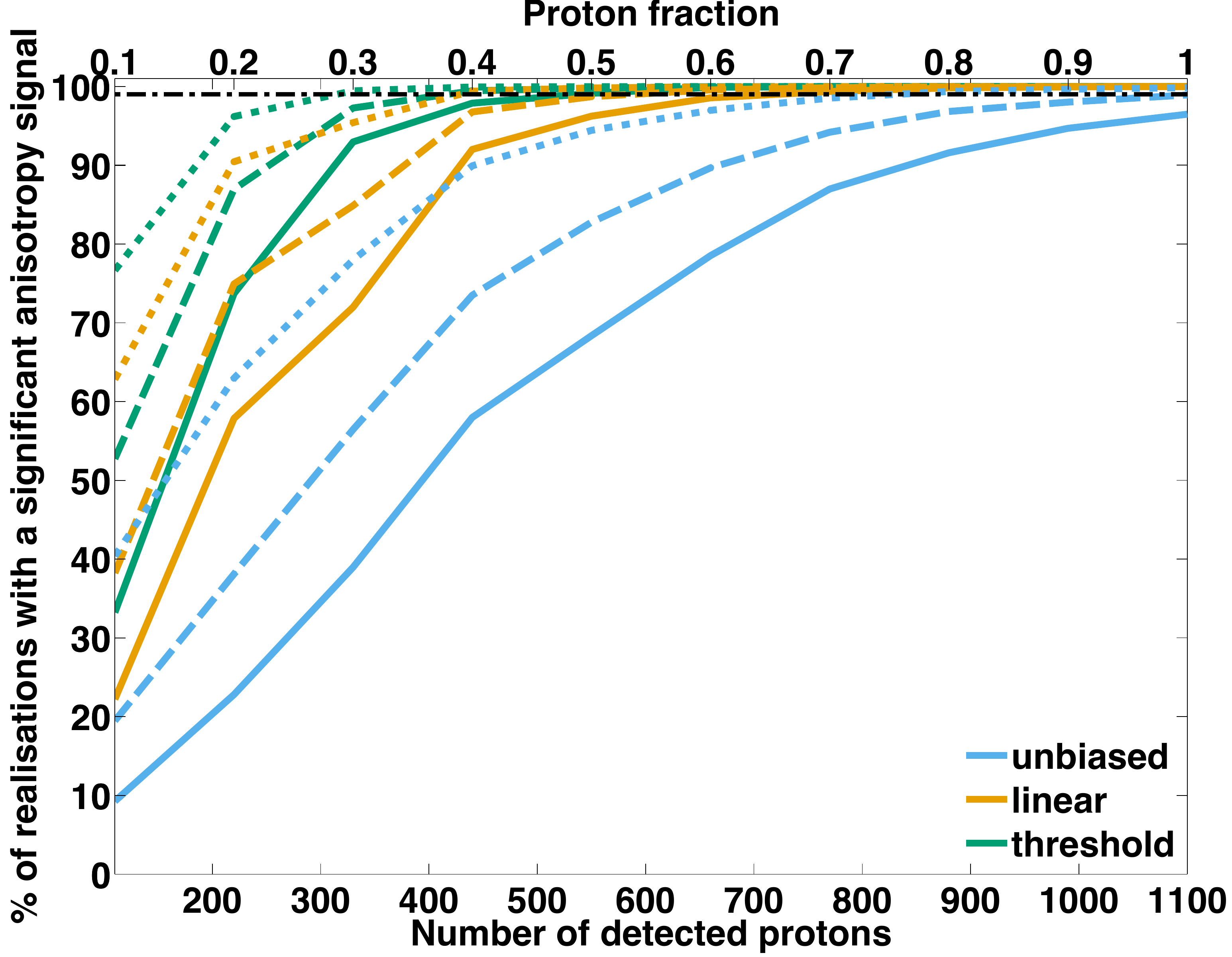}
\includegraphics[width=1.05 \figlength]{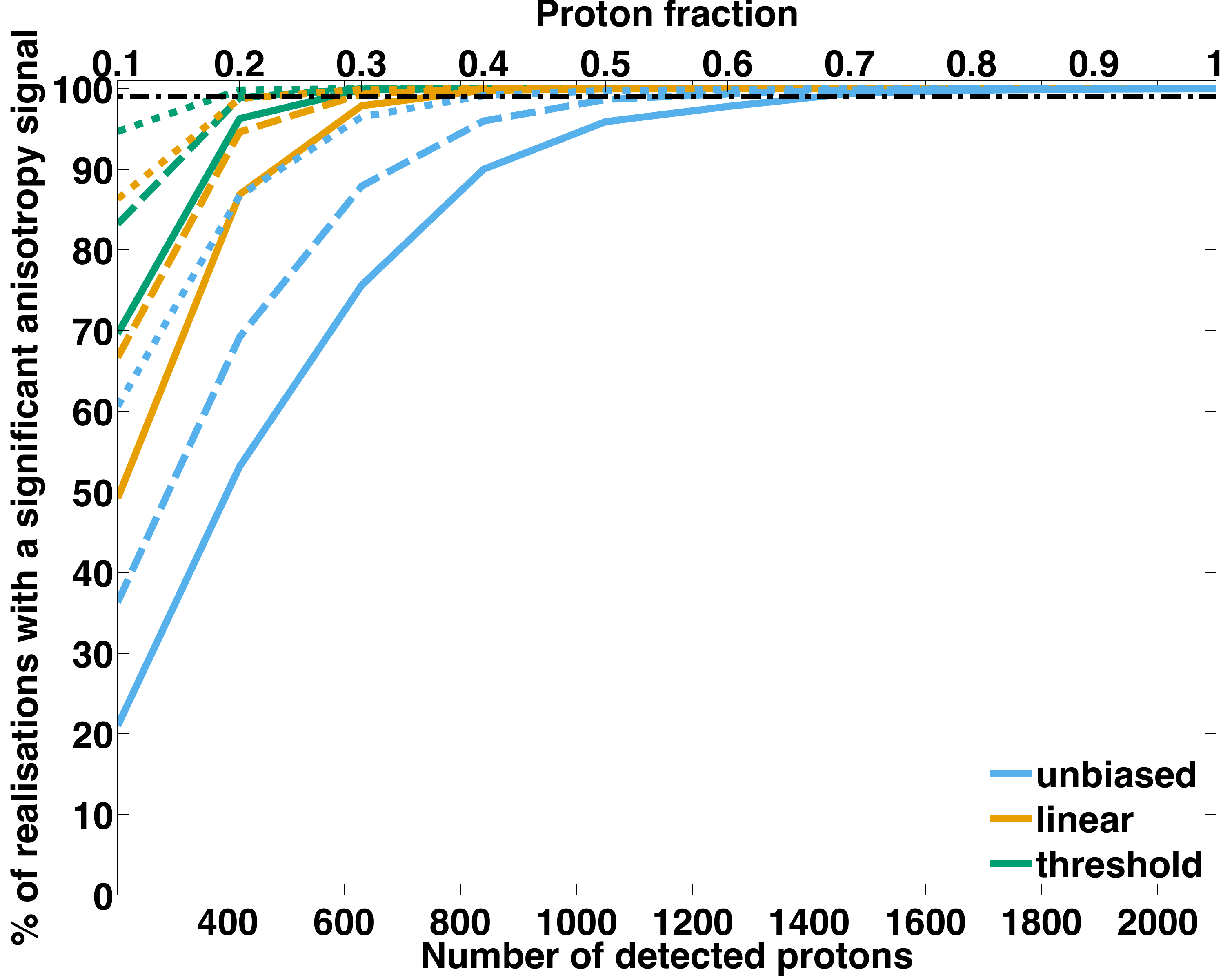}\\
\includegraphics[width=1.075 \figlength]{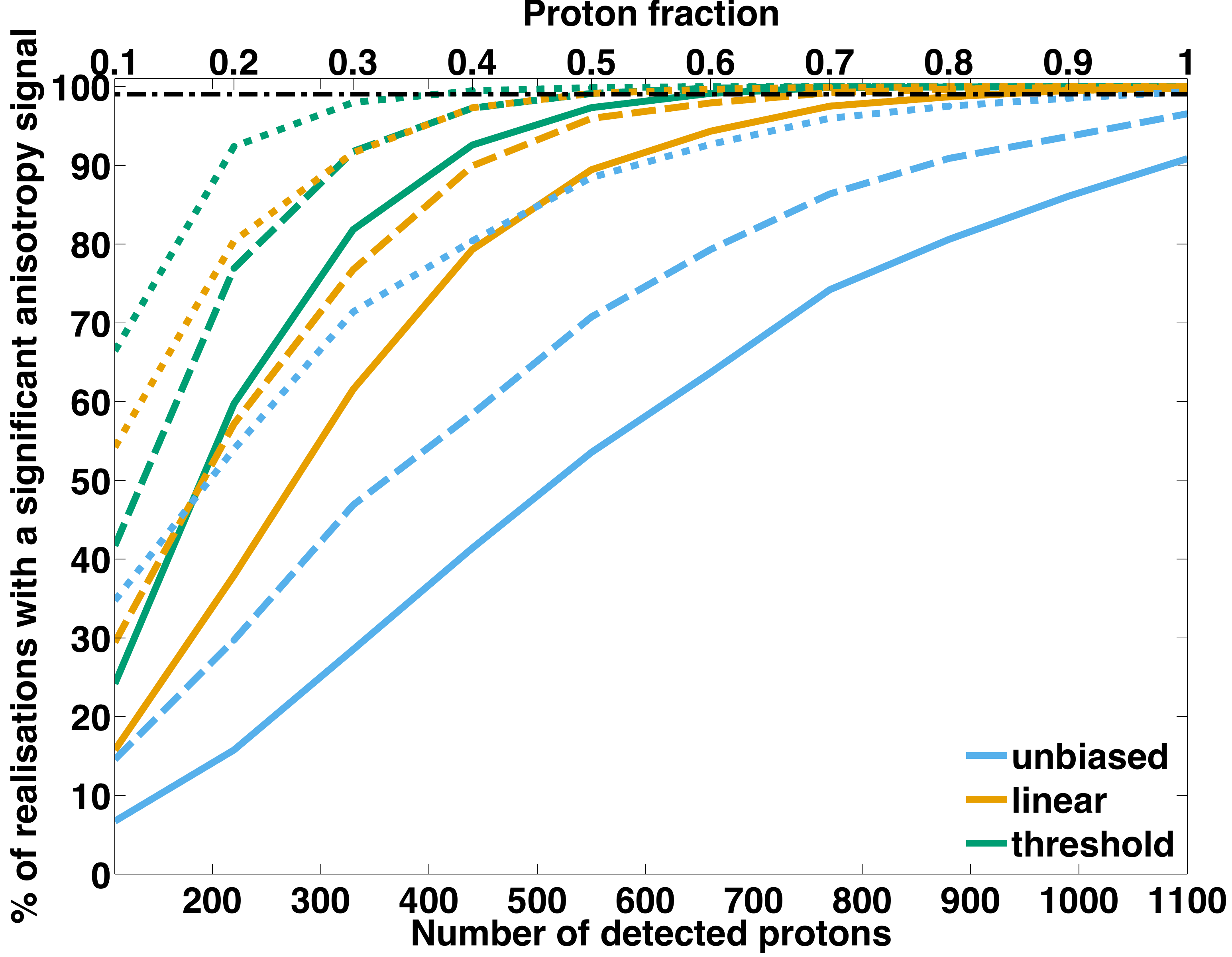}
\includegraphics[width=1.05 \figlength]{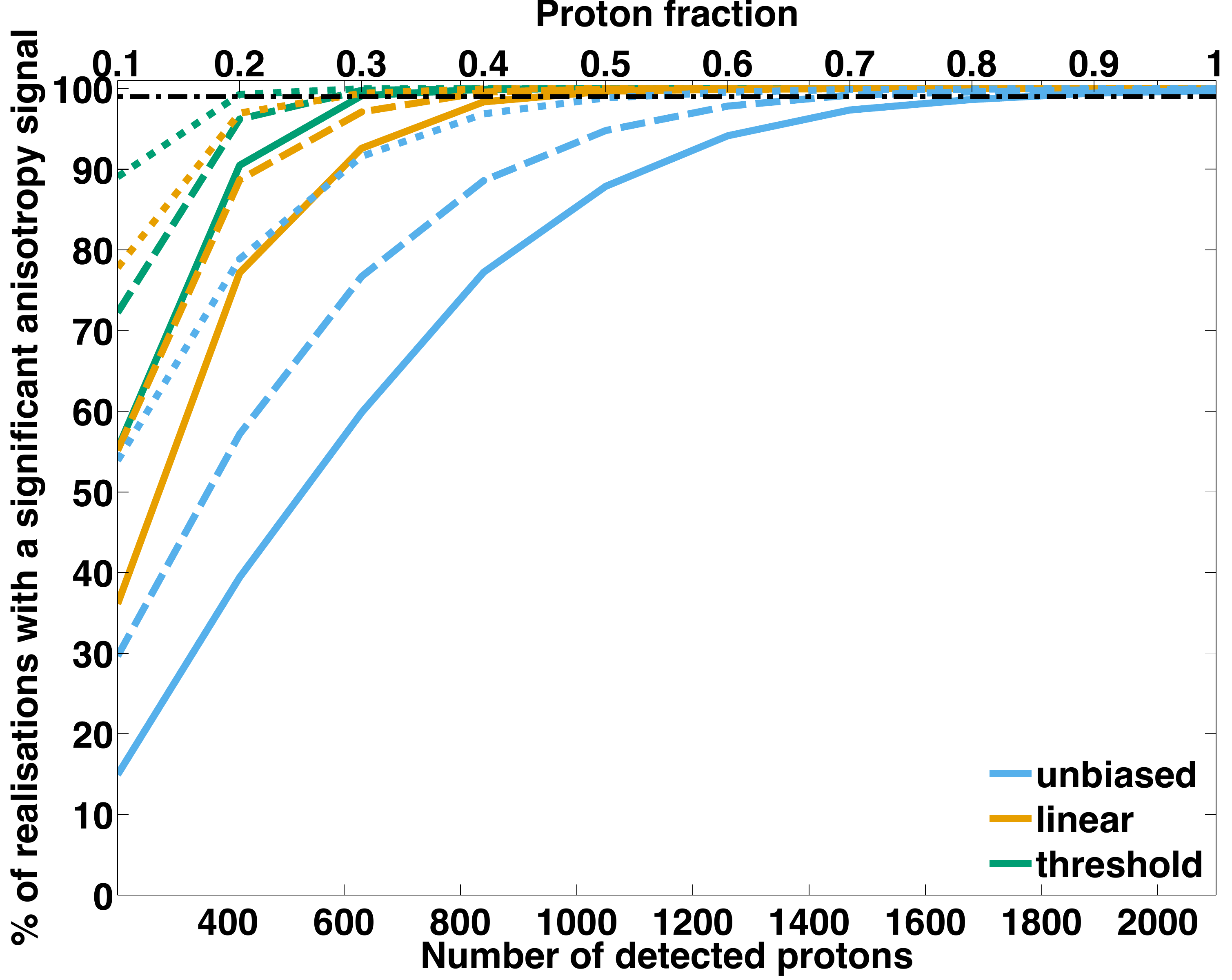}\\
\caption{The percentage of realisations of one of the models assumed for the UHECR source distribution (unbiased-light blue, linear bias-orange, threshold bias-green) in which an anisotropy is expected with significance $ \geq 95\%$ (dotted lines), $ \geq 99\%$ (dashed lines), $ \geq 99.7\%$ (solid lines) as a function of the number protons present in the data, assuming the composition of individual showers can be determined. The black dot-dashed horizontal line shows the $99\%$ CL. The UHECR source density is assumed to be $\bar{n} = 10^{-2}~{\rm Mpc}^{-3}$ on the top row and $\bar{n} = 10^{-3}~{\rm Mpc}^{-3}$ on the bottom row. Left (right) column panels assume 1100 (2100) events with $E \geq 50$ EeV will be detected in 5 years of JEM-EUSO.}
\label{fig:EXTRACTED_PROTONS}
\end{figure}

It is possible that the next generation UHECR experiment will be sufficiently sensitive to composition related observables, to determine the composition of the primary particles on an event-by-event basis, at least for a sub-sample of good quality events. In this section, we repeat the analysis of section \ref{subsec:bias}, assuming a fraction of light elements, present in the observed UHECR dataset, can be identified. In figure \ref{fig:EXTRACTED_PROTONS}, we plot the expected anisotropy signal in the various bias models considered, as a function of the number of observed UHECR protons with energy $E \geq 50$~EeV. The upper x-axis gives the fraction of the overall observed flux that the number of protons considered corresponds to, assuming a five-year JEM-EUSO exposure. We see that for a detection of anisotropy at the $\geq 95\%$ CL ($\geq 99.7\%$ CL) in the unbiased model, the number of observed protons should exceed 800 (1500), if $\bar{n} = 10^{-2}~{\rm Mpc}^{-3}$, or 1000 (1700) with an order of magnitude lower $\bar{n}$. A stronger anisotropy is expected in the linear bias model for the same number of observed protons, whereas in the threshold bias model an anisotropy at the $> 99.7 \%$ CL is expected, even with as few as 600 detected proton UHECRs.

\begin{figure}
\centering
\includegraphics[width=1.075 \figlength]{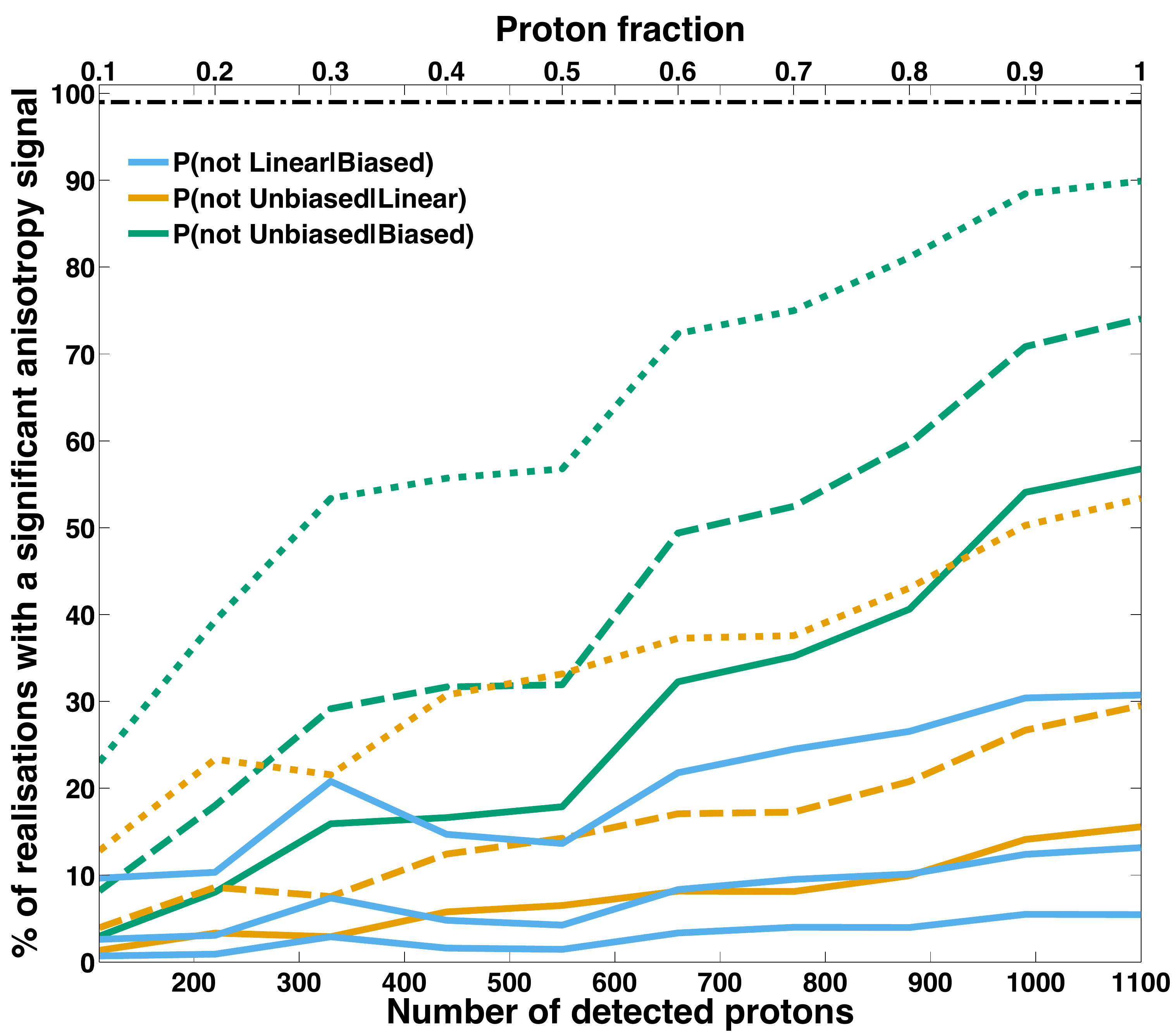}
\includegraphics[width=1.05 \figlength]{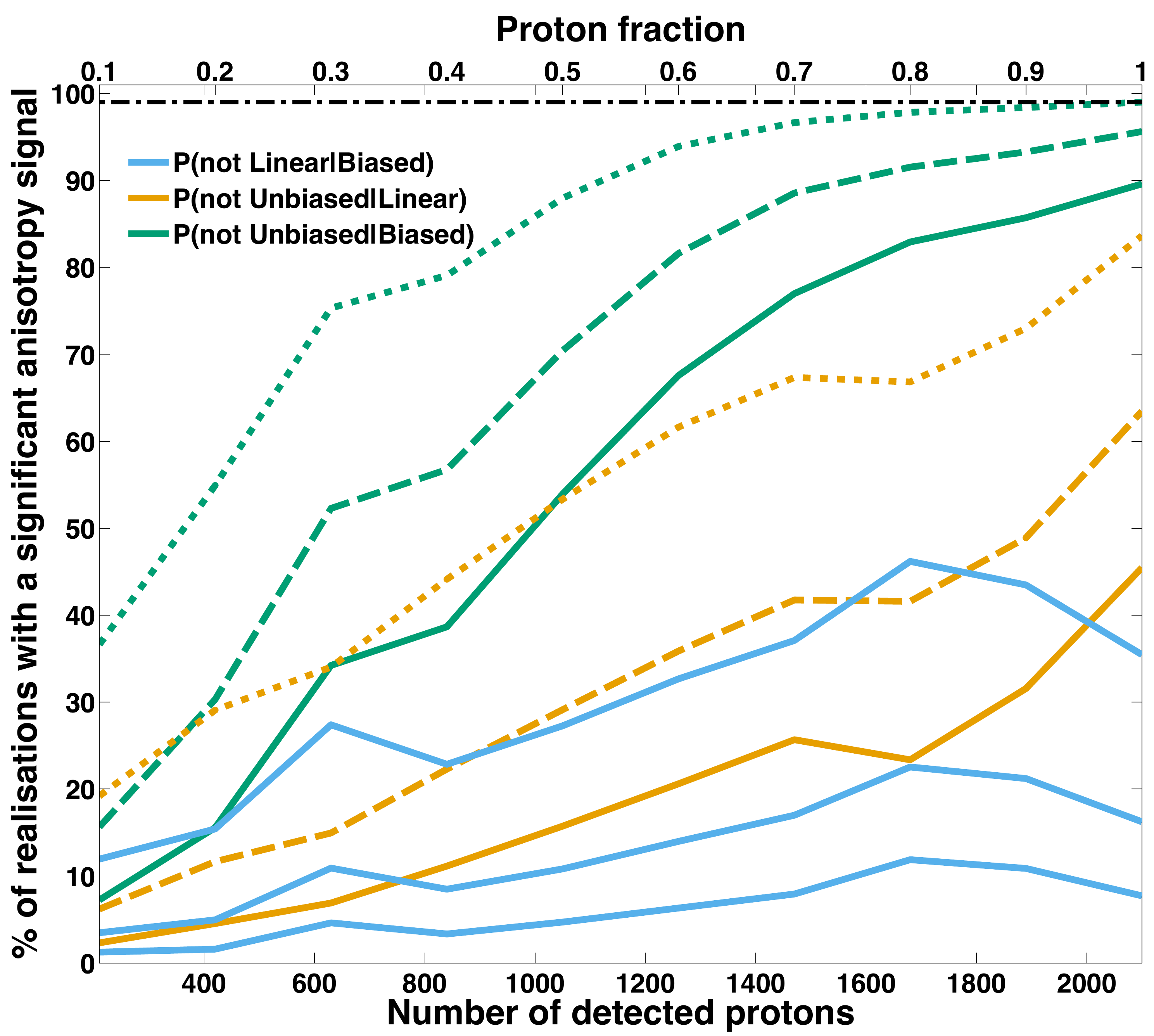}\\
\includegraphics[width=1.075 \figlength]{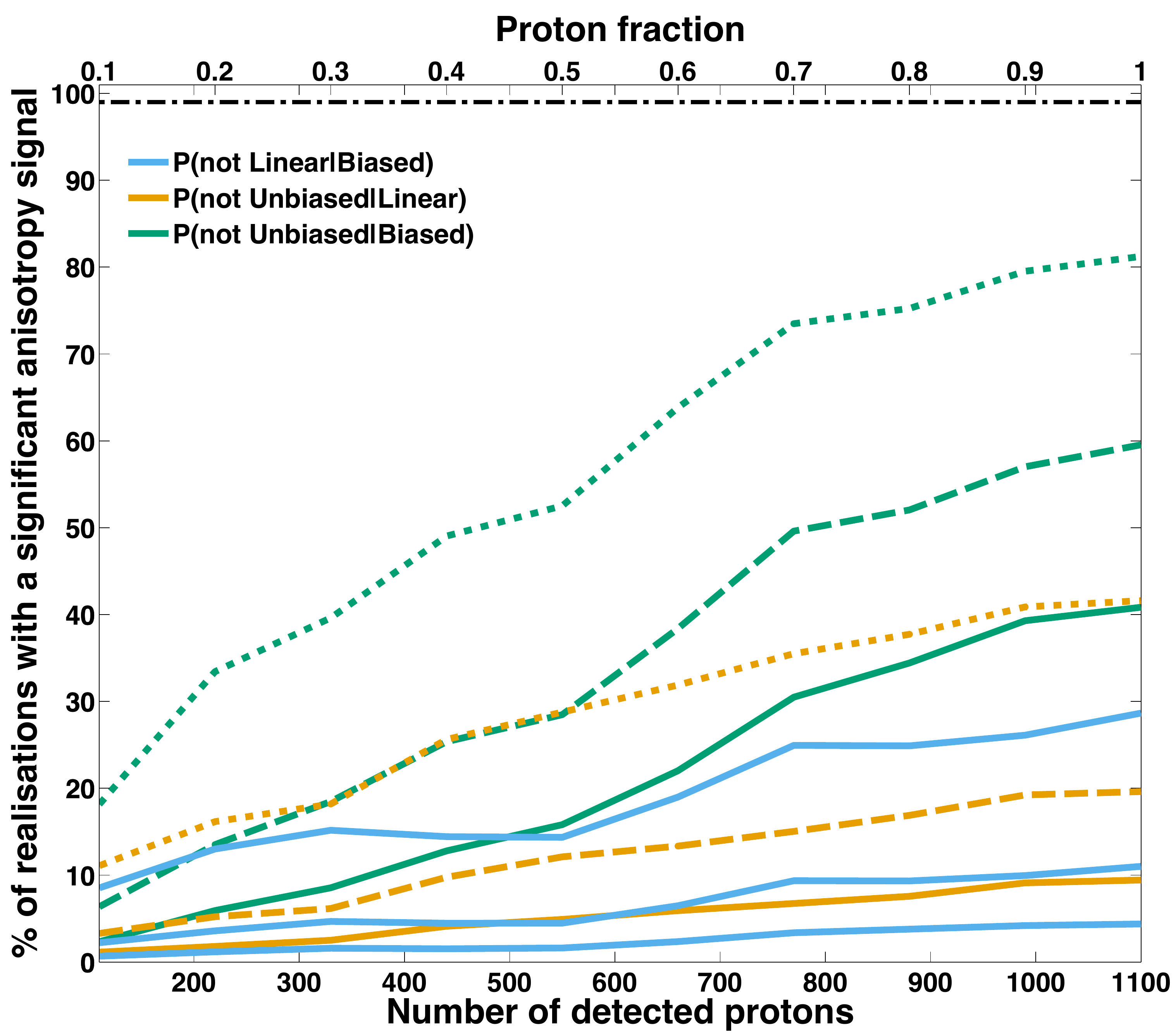}
\includegraphics[width=1.05 \figlength]{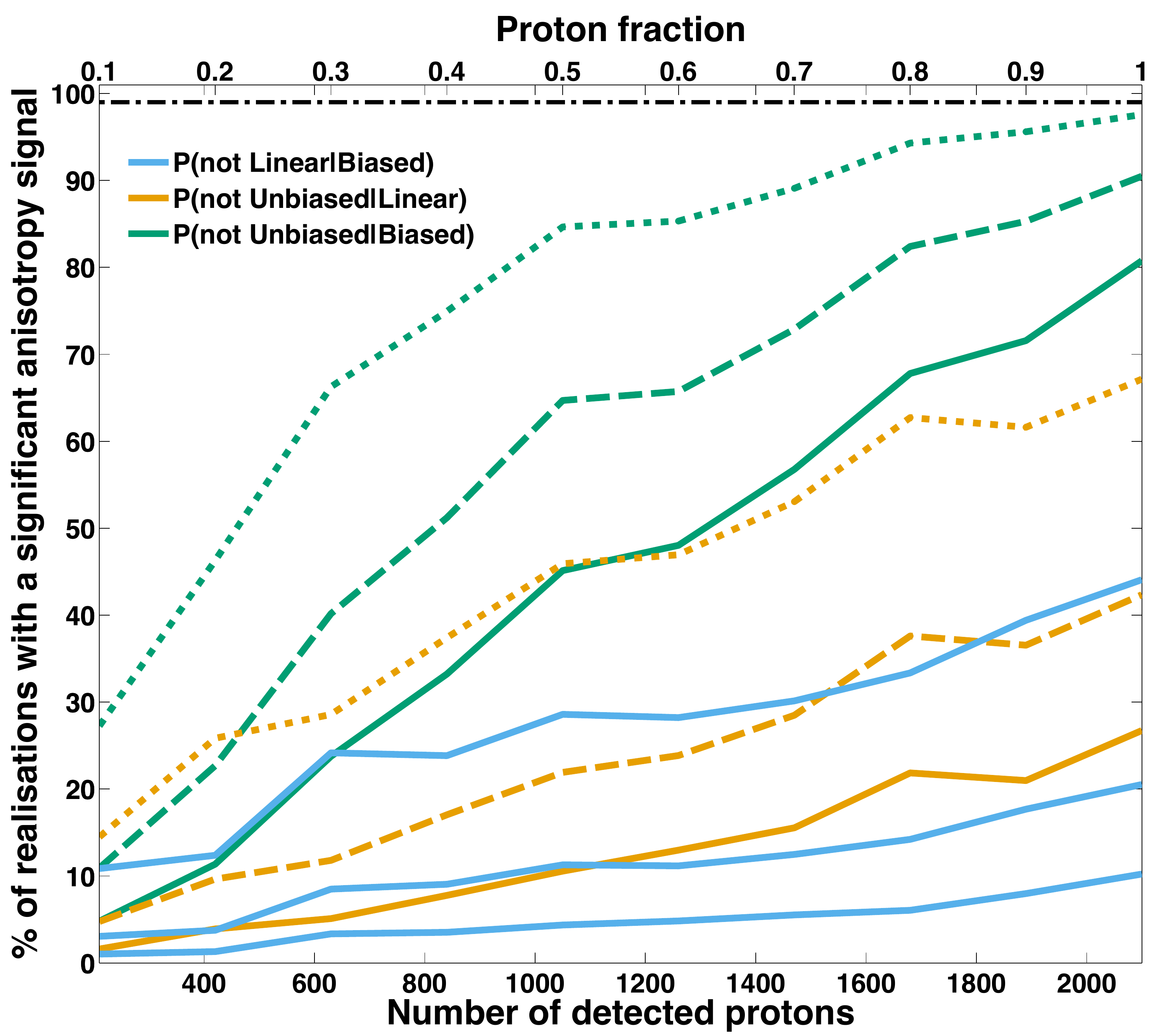}\\
\caption{Percentage of realisations of one of the models assumed for the UHECR source distribution, in which the probability of ruling out model $M_1$ assuming model $M_2$ is true P($\overline{M}_1 | M_2$), for the I-{\it isotropic}, UB-{\it unbiased}, L-{\it linear} and TH-{\it threshold} models, is $ \geq 95\%$ (dotted lines), $ \geq 99\%$ (dashed lines), $ \geq 99.7\%$ (solid lines), as a function of the number protons present in the data, assuming the composition of individual showers can be determined. The black dot-dashed horizontal line shows the $99\%$ CL. The UHECR source density is assumed to be $\bar{n} = 10^{-2}~{\rm Mpc}^{-3}$ on the top row and $\bar{n} = 10^{-3}~{\rm Mpc}^{-3}$ on the bottom row. Left (right) column panels assume 1100 (2100) events with $E \geq 50$ EeV will be detected in 5 years of JEM-EUSO.}
\label{fig:bias_models}
\end{figure}

Figure \ref{fig:bias_models} quantifies the probability of distinguishing between the unbiased, linear bias and threshold bias model considering the probability P($\bar{\rm M}_{1}| \rm M_{2}$) as above. We observe that a clear discrimination between a linear bias model and an unbiased model will not be possible even if $100\%$ of UHECRs are protons, for the conservative experimental resolution we have assumed in this work. A discrimination at the $95\%$ CL between the threshold bias model and the unbiased model is expected when the number of events beyond 50 EeV exceeds 2100, as long as the composition is proton dominated.

\section{Discussion}

We have studied the expected anisotropy signal in UHECR arrival directions which should be detectable with a future UHECR detector with an order of magnitude larger annual exposure to UHECRs at energy 100 EeV than Auger, under different astrophysical scenarios. Our results are general and apply to any such future full sky detector, although for definiteness we have assumed some of the characteristics of the proposed JEM-EUSO space telescope, which might be the next UHECR observatory to be realised. We have thus assumed a uniform full sky exposure, as well as the expected pointing, and energy resolution of JEM-EUSO, and its proposed detection efficiency. \\
\indent We constructed sky maps of the expected UHECR intensity in a range of models for the bias of UHECR sources relative to the galaxy distribution, motivated by the observed clustering of different astrophysical populations relative to the overall galaxy distribution. Motivated by recent measurements of UHECR composition, that suggest an increasingly heavy, mixed composition with energy above $\sim$20 EeV, we have conducted our analysis assuming a fraction of the observed UHECRs are deflected by large angles. We have simulated the effect of such heavily deflected UHECRs by assuming they arrive isotropically, smearing the expected anisotropy signal. For a given assumed fraction of protons in the data, this is almost certainly a conservative estimate of the expected anisotropy signal, as more than likely, at least some of the observed nuclei will retain some correlation with their sources, as shown for example in \cite{Rouille14}.\\
\indent We showed that if UHECR sources cluster to the matter distribution in a similar manner to that of galaxy clusters, a significant anisotropy should be detectable in the arrival directions when the number of events with energy $E\geq50$~EeV exceeds 2100 (1100) even for the modest, energy and pointing resolution of a space based instrument, as long as the fraction of protons at the highest energies is $\gtrsim 60\%$ ($80\%$). If proton like showers can be identified, anisotropy at the $3\sigma$ level is expected, in this model, once the number of detected protons exceeds 600, sensitive to the UHECR source number density. Further, we showed that if the UHECR source distribution follows the distribution of galaxies, with some linear bias comparable to the observed bias of low redshift AGN, then the expected anisotropy signal will be lower but detectable with $\gtrsim 3\sigma$ significance as long as the fraction of protons is $ \gtrsim 90\%$ in this energy range, and the UHECR source number density is not lower than $\bar{n} = 10^{-3}~{\rm Mpc}^{-3}$, assuming the energy scale of the Auger experiment. Assuming the energy scale of the TA experiment means that double the number of events should be detectable annually, and in that case an anisotropy should be expected at the $\gtrsim 3\sigma$ level, if the fraction of protons at the highest energies is $ \gtrsim 70\%$. We have pointed out the dependence of our results to the UHECR source number density throughout, and demonstrated that the larger the number of UHECR sources, the higher the expected anisotropy in this regime. In practice, the UHECR source number density should be possible to constrain with real data from the clustering in the dataset (number of ``repeaters'') as shown in \cite{Waxman95,kashti2008,Abreu:2013kif}. If the fraction of protons present in the dataset can be determined from composition observables, an anisotropy at the $\gtrsim 99.7\%$ CL is expected once the number of observed protons exceeds 1500, if the UHECR source distribution is $unbiased$ with respect to the galaxy distribution.\\
\indent Previously conducted simulation studies, referenced in section \ref{sec:intro}, have focused on the low source density regime and the highest cosmic-ray energies, emphasising the expectation that in this limit, due to the drastically smaller GZK horizon, an anisotropic arrival direction distribution is expected to be observed as a result of multiplets from individual ``bright'' UHECR sources each of which produces a significant fraction of the observed UHECRs above some energy cutoff (generally $E \gtrsim 80$~EeV). Instead we focused on the ``faint'' source regime, where the probability of multiplets from an individual source is low. In this latter regime, any anisotropy detected should be intrinsic to the clustering in the distribution of the sources. We demonstrated that even for a relatively low source number density, one should be able to probe the intrinsic anisotropy of the source distribution by considering lower energy events (as long as $E \gtrsim 40$~EeV, below which energy deflections are expected to be too severe, see e.g. discussion in \cite{kashti2008}). A lower limit to the source number density of order $\bar{n}\sim 10^{-5}\,{\rm Mpc}^{-3}$ has been derived from UHECR clustering in the Auger data \cite{Abreu:2013kif}.\\
\indent Throughout this work we have masked by the 2MRS mask leading, unfortunately, to an incomplete sky coverage. In some earlier anisotropy studies, the obscured region of the galactic plane and unobserved regions were populated by drawing the number of galaxies in the obscured part of the sky from a Gaussian, or Poisson distribution with a mean equal to the mean number of galaxies in adjacent observed regions. We have avoided using this technique as it can bias the results of the analysis.\\
\indent In conclusion, we have shown that an order of magnitude larger detector than current UHECR experiments will push forward the study of UHECRs. We have shown that a significant anisotropy should be expected in most astrophysical scenarios in the parameter space explored, consistent with the complementary findings of \cite{Rouille14}. An absence of anisotropy with such an increase in experimental exposure, could only mean a heavy dominated composition and/or unexpectedly large UHECR deflections, which could be translated to bounds on local magnetic fields or new physics. 

\section{Acknowledgements}
We thank P. Erdogdu for providing us with the 2MRS $K<11.75$ selection function. We thank O. Lahav, A. Olinto, E. Parizot, H. Takami and D. Waters for very fruitful discussions. FBA acknowledges the support of the Royal Society via a Royal Society, University Research Fellowship. KK thanks the Department of Physics and Astronomy of the University College London for its kind hospitality during this work. KK acknowledges financial support from PNHE and ILP. FO was supported by the Institut Lagrange de Paris (ILP) as a visitor at the Institut d'Astrophysique de Paris.

This work has made use of data from the Two Micron All Sky Survey, which is a joint project of the University of Massachusetts and the Infrared Processing and Analysis Center/California Institute of Technology, funded by the National Aeronautics and Space Administration and the National Science Foundation.
\label{sec:acknowledgements}

\bibliography{./OKA14}
\end{document}